\documentclass{article}
\usepackage{graphicx}
\usepackage{amsmath}
\usepackage{amssymb}
\usepackage{multicol}
\usepackage{etoolbox}
\usepackage{url}
\usepackage{float}
 \usepackage{booktabs}

\usepackage{xcolor}

\usepackage[left=0.9in,right=0.9in,top=0.9in,bottom=0.9in]{geometry}

\usepackage[numbers,sort&compress]{natbib}

\usepackage{setspace}
\usepackage{authblk}

\usepackage{lineno}

\title{Scaling intra-urban climate fluctuations}
\author[1]{Marc Duran-Sala}
\author[1]{Martin Hendrick}
\author[1,*]{Gabriele Manoli}
\affil[1]{Laboratory of Urban and Environmental Systems, École Polytechnique Fédérale de Lausanne, Switzerland}
\affil[*]{Corresponding author: gabriele.manoli@epfl.ch}
\date{\today}

\begin{document}


\maketitle

\noindent\textbf{Published version:} Nature Cities (2026). https://doi.org/10.1038/s44284-026-00441-z

\begin{abstract}
Urban-induced microclimate variations, such as urban heat islands and air pollution, scale with city size, producing distinctive relations between average climate variables and city-scale quantities (e.g., total population). However, these relations are sensitive to city boundary definitions and overlook intra-urban variability. Here, we overcome these limitations by using high-resolution data of urban temperatures, air quality, population, and street networks from 142 cities worldwide, showing that their marginal and joint probability distributions collapse onto a set of general functions inspired by finite-size scaling in statistical physics. Through a logarithmic relation linking urban spatial features to climate variables, we find that average street network properties are sufficient to characterize the full variability of temperature and air pollution fields within and across cities. These findings show that intra-urban climate variability follows general scaling functions, enabling the integration of climate information into reduced-complexity models of urban systems to better inform future urban planning.

\end{abstract}

\section*{Introduction}
High temperature and air pollution in cities increase the risk of heat- and pollution-related mortality, posing serious economic and public health challenges \cite{Huang2023,Estrada2017,khomenko2021}. In parallel, urban areas are growing, with projections indicating that 68\% of the global population will reside in cities by 2050 \cite{UN2018}. Understanding how urbanization impacts local climate is therefore critical to reduce risks and shape healthier urban environments \cite{Chapman2017, Wang2019}.

While the physical processes behind urban-induced changes — such as the Urban Heat Island (UHI) effect and the increase in Urban Particulate Matter (UPM) concentrations — are well understood, existing research has mostly focused on city-specific studies based on detailed monitoring and modeling efforts \cite[e.g.,][]{Oke2017}. This has substantially advanced our understanding of mass and energy transfers in urban contexts, including the impacts of different building materials, urban geometries, and human activities on local microclimate \cite[e.g.][]{Oke2017}, but the identification of general patterns and laws that are valid across diverse cities and geographic contexts is lagging behind \cite{Lobo2023, Manoli2025, Creutzig2025}. Urban complexity research has emerged as a powerful approach to explore such emergent behaviors \cite{Batty2009,Bettencourt2021,Caldarelli2023}, providing the methods and theoretical frameworks that are needed to describe the structure and dynamics of cities in the most general terms. Yet, translating this knowledge into a general description of urban microclimate remains an open challenge. 

\begin{figure}[H]   
    \centering
    \includegraphics[width=\linewidth]{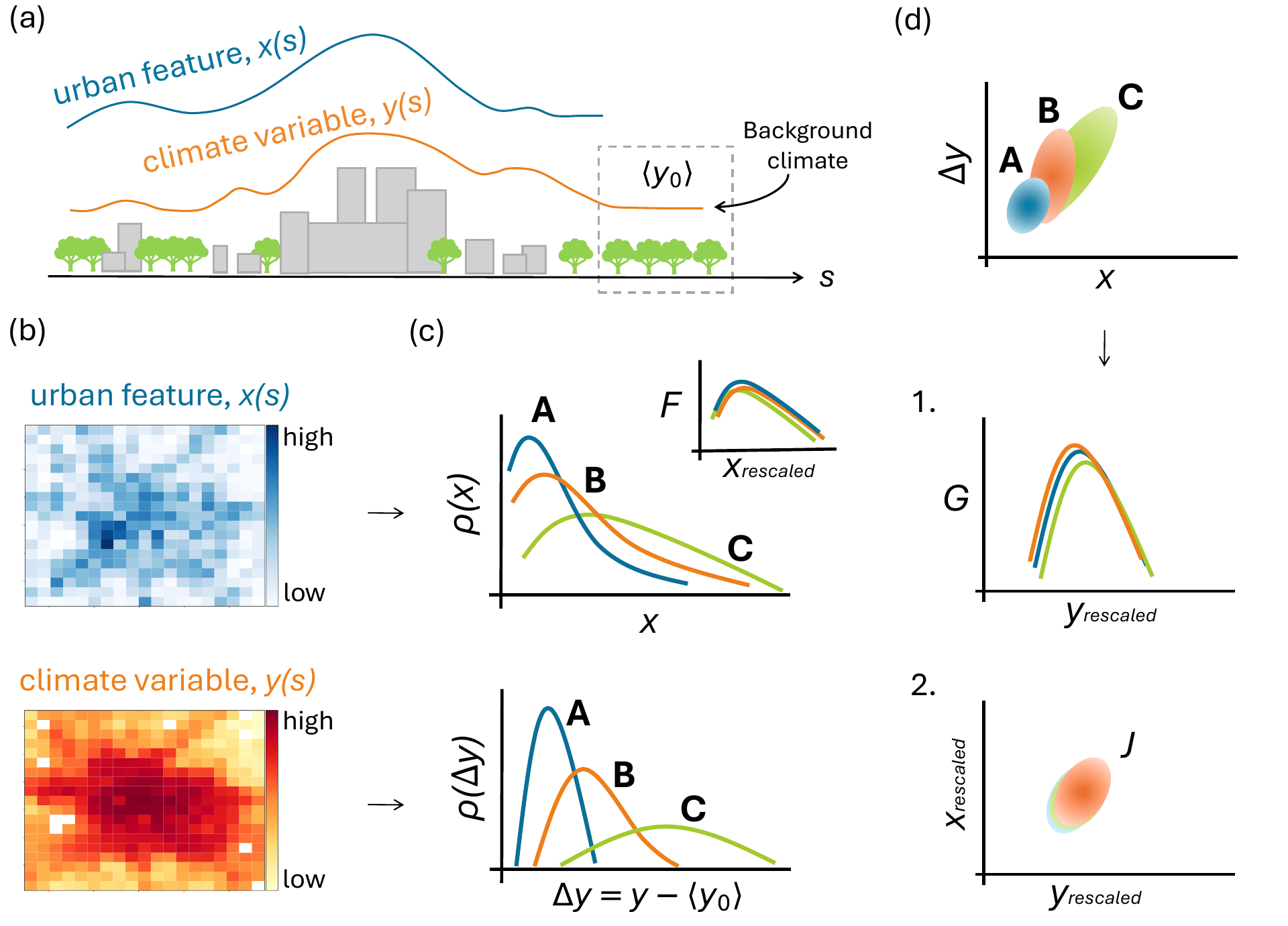}
    \caption{\footnotesize 
    \textbf{Conceptual framework for analyzing the covariation of urban structure and climate.} 
    (a) In this study we consider intra-urban variations of climate variables $y(s)$ (i.e., temperature $T$ and particulate matter concentrations $PM$) and their relation to urban features $x(s)$ (i.e., street intersections $n$ and population counts $p$), where $s$ is the spatial coordinate. Urban-rural differences $\Delta y = y-\langle y_{0}\rangle$ are calculated considering background climatic conditions $\langle y_{0}\rangle$.
    (b) Example of  a regular square grid  used to aggregate all datasets, allowing both climate variables and urban features to be compared across cities.
    (c) PDFs of urban features $\rho(x)$ and urban climate perturbations $\rho(\Delta y)$, where each curve represents one city. Previous work \cite{Hendrick2024} showed that the PDFs of urban features collapse onto a single curve following a finite-size scaling form $F$ (inset).
    (d) Methodology employed in this study: first we investigate the relation between $x$ and $\Delta y$ and then we rescale their marginal and joint PDFs to obtain a data collapse - indicating the existence of general scaling functions ($G$ and $J$, see main text).}
\label{fig:1}
\end{figure}

Consider an urban area characterized by some properties $x$ (e.g., population) that are heterogeneous in space, that is $x=x(s)$, where $s$ is the spatial coordinate (Fig. \ref{fig:1}a). Such urban characteristics modify local climate variables $y$ (e.g., temperature) by perturbing the background climatic conditions $\langle y_0 \rangle$, where the brackets $\langle .\rangle$ indicate spatial averaging. This induces a distinct field of urban-rural differences $\Delta y(s) = y(s)  - \langle y_0 \rangle$, which are often aggregated at the city level to define quantities such as the UHI intensity \cite[e.g.,][]{Manoli2019}, that is $\langle \Delta y  \rangle = \frac{1}{A} \int_{A} y(s)ds - \langle y_0 \rangle$, where $A$ is the urban area. Drawing parallels to allometric scaling in biological systems \cite[e.g.,][]{West2005, kleiber1932}, previous studies have applied scaling laws to quantify how these aggregate climate variables scale with city size i.e., $\langle \Delta y  \rangle \propto X^b$, where the climate perturbation $\langle \Delta y  \rangle$ is defined in terms of air temperature \cite{Li2020,Sobstyl2018}, surface temperature \cite{Zhao2014,Manoli2019}, or air pollution \cite{Han2014,Lamsal2013}, $b$ is the scaling exponent, and $X=\int_{A} x(s)ds$ is a given city-scale urban feature generally defined by total urban population, urban area, or gross building volume \cite[e.g.,][]{Manoli2019,Li2020}.

However, unlike biological systems, cities lack clear perimeters and the construction of scaling laws depends on the choice of the area $A$, thus inevitably introducing biases related to boundary definitions. Previous studies \cite{Cottineau2017, Arcaute2015} have shown that even small variations in density thresholds, commuting flows, or population cut-offs used to delineate urban areas, lead to different scaling exponents $b$. Beyond these parametric approaches, the choice between administrative versus metropolitan boundaries can also yield opposite conclusions — for example, on whether larger cities emit more or less CO$_2$ per capita (i.e., $b>1$ versus $b<1$) \cite{Louf2014}. Moreover, urban systems can be highly heterogeneous in terms of infrastructure and population densities \cite[e.g.,][]{Volpati2018}, with local climatic conditions varying from building to neighborhood scales due to diverse materials, green spaces, urban geometries, and anthropogenic activities \cite[e.g.,][]{Vardoulakis2011,Qiu2017}.

With regards to urban features $x$, these limitations have been recently overcome by Hendrick et al. \cite{Hendrick2024}, who — inspired by analogous findings on biological metabolism \cite{Giometto2013,Zaoli2017,Zaoli2019,Botte2021} — demonstrated that the entire variability of population and street intersections within a city, as described by their probability density functions $\rho(x)$, follows a finite-size scaling (FSS) form:
\begin{equation}
  \rho(x \mid \langle x \rangle) = x^{-\lambda} \, F \left( \frac{x}{\langle x \rangle^\delta} \right),
  \label{eq:hendrick}
\end{equation}
where $F$ is a scaling function providing a finite-size cutoff, and $\lambda$ and $\delta$ are the scaling exponents that describe how the distributions change with system size and provide a physical interpretation of the characteristic scale. The two exponents are not independent due to mathematical constraints (i.e., normalization). When $\delta = 1$, the distributions have the same shape and differ only in their characteristic size, whereas for $\delta \neq 1$ their shape varies systematically. Therefore, the value of the scaling exponent determines the physical meaning of the scale parameter (i.e., $\langle x \rangle^{\delta}$ in equation \eqref{eq:hendrick}). In this framework, since $\lambda \approx \delta \approx 1$, the scale corresponds to the mean value $\langle x \rangle$, which is therefore sufficient to characterize the full probability distribution of urban features $\rho(x)$ — i.e., using equation \eqref{eq:hendrick}, distributions from cities around the world collapse onto a single master curve $F$, which is approximately log-normal \cite{Hendrick2024}.
More broadly, this approach is rooted in the FSS framework of statistical physics \cite{fisher1972scaling,cardy2012finite}, where "scaling laws" observed in finite systems are understood as effective manifestations of an underlying scale-invariant behavior that is regularized by finite size. In FSS, one does not expect pure power laws to hold unconditionally; instead, system-size dependence is absorbed into a characteristic scale, and the residual variability is captured by a universal scaling function — revealed empirically through data collapse onto a master curve. Equation \eqref{eq:hendrick} is precisely of this form: the exponent $\lambda$ controls the apparent scaling regime, while $\langle x\rangle^\delta$ plays the role of the size-dependent cutoff that encodes how the finite urban “system” constrains fluctuations. Since cities are finite in size, interpreting them through the lens of FSS naturally reframes allometric relations as "effective" scaling behaviors, and provides a principled route to compare heterogeneous cities by separating universal structures (the master curve) from size-, morphology-, and climate-dependent corrections to scaling.

The FSS framework serves here as a point of departure to explore whether a similar ansatz exists also for urban climate (see Fig.~\ref{fig:1} for an overview of our conceptual framework and methodology). Because cities differ widely in size, morphology, and background climate, their temperature and pollution distributions are not directly comparable. Thus, we investigate whether the probability distributions of climate variables $\rho(y)$ follow a general scaling form. In addition, given that aggregate urban climate statistics are generally unknown for cities around the world, we ask whether information on urban structure $x$ is sufficient to characterize the entire intra-urban variability of climate (defined here as mean annual temperature and air pollution fields).

The description of the spatial variability of climatic conditions has long been a central focus in urban climatology \cite[e.g.,][]{Oke2017}. Yet, a general statistical framework describing intra-urban climate variations — and its link to urban form and characteristics — is still missing. An attempt in this direction is represented by radial decay models \cite{Zhou2015,Yu2020} which relate urban-induced microclimate changes to the distance from the city center. These approaches indicate that both temperature and pollution levels follow a decreasing trend with radial distance, but they focus only on the mean behavior of climatic characteristics, thus neglecting the spatial heterogeneities of cities — which can markedly deviate from average coarse-grained quantities. Here, we overcome all the aforementioned limitations by showing that marginal probability distributions of urban climate — and their joint distributions with urban features — are characterized by general scaling functions which, upon proper rescaling, collapse onto a unique curve for each climate variable, reflecting a common statistical form shared across cities. These results provide a novel stochastic description of intra-urban climate variability that is valid across cities, generalizes previous findings on urban structure \cite{Hendrick2024}, and improves traditional urban climate decay models \cite[e.g.,][]{Zhou2015}. Given the climate challenges faced by cities worldwide, this knowledge is critical to assess the environmental exposure of urban residents and inform climate-sensitive urban planning strategies.

\section*{Results}\label{sec:results}

Our analysis investigates intra-urban climate variability using high-resolution datasets from 142 cities worldwide — ranging from small towns with tens of thousands of inhabitants (e.g., Tartu, Estonia) to megacities with populations exceeding one million (e.g., Bogotá, Colombia), spanning all major climate zones and levels of urban development across the world (see Supporting Information, SI Table~\ref{table:SI_cities} and Fig.~\ref{fig:2}c). Specifically, we analyze near-surface air temperature ($T$) and particulate matter concentrations ($PM_{2.5}$, denoted as $PM$ for simplicity) — which represent the generic climate variable $y$ — alongside the number of street intersections ($n$) and population counts ($p$) — which represent the urban features $x$ (see Methods for details).

To ensure consistent spatial resolution across cities, we discretize space into a regular grid of square cells of size $l = 1000~m$, aggregating all the data within each cell (Fig.~\ref{fig:1}b; see Methods). Thus, we obtain pairs $(x_i,y_i)$ for each grid cell $i$, where $x_i$ is the urban feature (e.g., street intersections) and $y_i$ is the climate variable (e.g., temperature). For urban climate, we define urban and non-urban grid cells and calculate urban-rural differences for each urban grid cell $i$ as $\Delta y_i = y_i - \langle y_0 \rangle$, where $\langle y_0 \rangle$ is computed as the average value of the non-urban grid cells surrounding each city (see Methods for details).

\subsection*{Scaling intra-urban climate variations across cities}
\label{sec:climate_scaling}
While distributions of urban features are strongly asymmetric and can be fully characterized by the mean \cite{Hendrick2024}, climate distributions appear shifted and nearly symmetric around the mean (see Fig.~\ref{fig:2}a and Fig.~\ref{fig:3}a for temperature and pollution, respectively). Hence, for climate variables we propose a location-scale ansatz that differs from equation \eqref{eq:hendrick} as it requires two rather than one parameter — the location $\langle \Delta y \rangle$ and the scale parameter $\sigma_{\Delta y}^\gamma$. Because cities differ in their background climate and in how strongly urban form perturbs that field, we (i) remove each city’s rural baseline to isolate the urban contribution and (ii) locate and scale each PDF by a characteristic amplitude of intra-urban climate intensity and fluctuations. In particular, we found that the probability density functions (PDFs) of the climate variables follow a scaling ansatz of the form:
\begin{equation}
    \rho(y | \langle \Delta y \rangle , \sigma_{\Delta y}) = \frac{1}{\sigma_{\Delta y}^{\gamma}} G \big( y_{\mathrm{rescaled}} \big),
    \label{eq:ansatz}
\end{equation}
where $G$ is a general scaling function that holds across all cities, the rescaled climate variable is defined as
\begin{equation}
    y_{\mathrm{rescaled}} =
    \frac{\,y - \langle y_0 \rangle - \langle \Delta y \rangle}
         {\sigma_{\Delta y}^{\phi}},
    \label{eq:y_rescaled}
\end{equation}
$\phi$ and $\gamma$ are the scaling exponents, and $\langle y_0 \rangle$ is the rural mean of temperature or PM concentrations (i.e., $\langle T_0 \rangle$ and $\langle {PM}_0 \rangle$, respectively).
This approach allows to collapse the different PDFs into a single curve $G$ (both for $T$ and $PM$), revealing that the intra-urban climate fields can be described by a general statistical form across cities (see Fig.~\ref{fig:2}b and Fig.~\ref{fig:3}b).

\subsection*{Linking climate perturbations to urban features}
\label{sec:climate_features_link}
The result in equation \eqref{eq:ansatz}, while notable for its validity across all the analyzed cities, relies on a priori knowledge of urban climate variables (i.e., $\langle \Delta y \rangle$ and $\sigma_{\Delta y}$), which are generally unknown. Hence, as a next step, we seek to predict the climate PDFs in equation \eqref{eq:ansatz} using information on urban features rather than empirical statistics of $T$ and $PM$. To do so, we propose a simple model that describes their covariation within cities. Specifically, we found that urban-rural differences in temperature $\Delta T$ (i.e., the magnitude of UHI) and particulate matter concentration $\Delta PM$ are well captured by a logarithmic relationship with either population counts or the number of street intersections (see SI Fig.~\ref{fig:SI_log_results}). Explicitly, for the grid cell $i$ in a given city, this relationship is expressed as:
\begin{equation}
    \Delta y_i = \alpha + \beta \thinspace \ln x_i,
    \label{eq:log}
\end{equation}
where $\alpha$ and $\beta$ are city dependent intercept and slope, respectively. The analysis of the $\alpha$ and $\beta$ values across all the cities analyzed reveals that their magnitude is more constrained when derived from street intersections compared to population counts (see SI Fig.~\ref{fig:SI_log_results}). Note that equation \eqref{eq:log} is statistically appropriate because $T$ and $PM$ are approximately normally distributed (Fig.~\ref{fig:2}a and Fig.~\ref{fig:3}a), while urban characteristics follow approximately log-normal distributions \cite{Hendrick2024}. For the sake of completeness, we also performed a numerical comparison with linear and power-law models and discussed a possible multivariate extension of the logarithmic model (for further details see SI Table~\ref{table:SI_fit_models} and SI Section~\ref{sec:multiloglinear}). However, our primary goal is not to predict fine-scale climate values in each square cell $i$ (as typically done by regression models; e.g., \cite{Briggs1997}), but rather to predict aggregate statistical properties of climate field — i.e., the location and scale. Thus, applying the expected value and variance to both sides of equation \eqref{eq:log} yields the following expressions for the mean and standard deviation of climate variables:
\begin{equation}
    \langle \Delta y (x) \rangle = \alpha + \beta  \langle \ln x \rangle \quad \text{,} \quad \sigma_{\Delta y} (x) = \beta \sigma_{\ln x}.
    \label{eq:log_mean_std}
\end{equation}
which provide a link between quantifiable urban feature information and generally unknown urban climate variables. Thanks to the relations in equation \eqref{eq:log_mean_std}, we can now estimate the location and scale parameter in equation \eqref{eq:ansatz} using the statistical properties of urban features and reconstruct the distributions of urban climate variables without any a priori information on their statistics.

To reduce the number of free parameters while retaining generality, we replace city-specific intercepts $\alpha$ and slopes $\beta$ with a small set of cluster-specific averages, $\overline{\alpha}_k$ and $\overline{\beta}_k$, where $k \in \{1,\ldots,K\}$ indexes clusters and $K$ is the total number of clusters. Intuitively, clustering allows cities that have similar environmental contexts (regional climate, local emissions regulations or topography) to share parameters, yielding a parsimonious model with $2K$ parameters instead of $2\times(\text{number of cities})$.
Thus, with cluster-specific parameters, equation \eqref{eq:ansatz} becomes:
\begin{equation}
    \rho(y | \langle \ln x \rangle, \sigma_{\ln x}) \cdot \,(\overline{\beta}_k \sigma_{\ln x})^{\gamma} = G\big(y_{\mathrm{rescaled}}(x)\big),
    \label{eq:ansatz_rescaled_kmeans}
\end{equation}
where the rescaled climate variable is now defined as:
\begin{equation}
    y_{\mathrm{rescaled}}(x) =
    \frac{\,y - \langle y_0 \rangle - \big(\overline{\alpha}_k + \overline{\beta}_k \langle \ln x \rangle\big)\,}
         {(\overline{\beta}_k \sigma_{\ln x})^{\phi}},
    \label{eq:y_rescaled_kmeans}
\end{equation}
where $x$ denotes the chosen urban feature (street intersections $n$ or population counts $p$).

We first tested a single pair of global parameters ($K{=}1$), i.e., intercept $\overline{\alpha}$ and slope $\overline{\beta}$ common to all 142 cities. The resulting data collapse for both temperature and $PM$ revealed distinct families of curves (SI Fig.~\ref{fig:SI_collapse_clustering}), indicating that environmental context matters in addition to urban features, which we quantified using the k-means clustering method to group cities with shared parameter values $\overline{\alpha}_k$ and $\overline{\beta}_k $(see Methods for details). While both street intersections and population can serve as predictors for rescaling the urban climate PDFs, Fig.~\ref{fig:SI_clustering_kmeans_popu_vs_node} shows that, for temperature, fewer clusters $K$ are needed when using street intersections, whereas for pollution, both predictors perform similarly. Consequently, we adopt the street network ($x{=}n$) as our predictor, for which the optimal partition is $K{=}3$ (Fig.~\ref{fig:2}c for temperature and Fig.~\ref{fig:3}c for $PM$; see SI Section~\ref{sec:clustering_effect} and SI Table~\ref{table:SI_cluster_alpha_beta_absbeta} for details on cluster characteristics).

After rescaling via equations \eqref{eq:ansatz_rescaled_kmeans}–\eqref{eq:y_rescaled_kmeans}, all PDFs collapse onto a common curve (Fig.~\ref{fig:2}d and Fig.~\ref{fig:3}d), indicating that, once structural differences are accounted for, intra-urban climate variability follows a common statistical form — consistent with evidence that local anomalies are modulated by urban morphology and form \cite{Sobstyl2018,Li2020}. For details on the rationale behind the clustering effect on the data collapse we refer readers to SI Section~\ref{sec:clustering_effect}.

To identify the scaling exponents that yield the best data collapse, we employed a residual metric $R$ \cite{Bhattacharjee2001}, which measures the cumulative area enclosed between all pairs of rescaled PDF curves. This provides a quantitative criterion for the quality of the data collapse. A lower $R$ indicates better collapse, reflecting that the rescaled curves are closer together. The scaling exponents are not independent due to the normalization constraint \(\int_{0}^{\infty} \rho(y)\,dy = 1\) ($\phi=\gamma$, see SI Section \ref{sec:scaling_exponents}). Thus, minimizing $R$ over $\phi$, shows that the best collapse occurs for $\phi=\gamma \approx 1$ (insets in Fig.~\ref{fig:2}d and Fig.~\ref{fig:3}d). This result — consistent with the proportionality of moments derivations — implies that the distributions preserve their shape across cities and differ only in their characteristic scale, which corresponds to the standard deviation — confirming that intra-urban climate distributions are fully characterized by their mean and variance (see SI Section \ref{sec:scaling_exponents} and SI Fig.~\ref{fig:SI_mean_var}).

Note that the functional form of the collapsed curves are reasonably well approximated by a normal distribution (dashed black line in Fig.~\ref{fig:2}b-d and Fig.~\ref{fig:3}b-d). As a matter of fact, equation \eqref{eq:ansatz} can be derived using the results by Hendrick et al. \citep{Hendrick2024} for urban features — i.e., equation \eqref{eq:hendrick} — together with the logarithmic relation in equation \eqref{eq:log}, leading to a Gaussian form of $G$ (see derivation in the SI Section~\ref{sec:urban_climate_universal_scaling_form}).

\subsection*{Covariation of intra-urban characteristics and climate variables}

After showing that the PDFs of temperature and PM concentrations follow the scaling ansatz in equation \eqref{eq:ansatz}, we now investigate if similar arguments apply to their joint probability distributions with urban features (Fig.~\ref{fig:2}e and Fig.~\ref{fig:3}e). Since urban climate variability is influenced by multiple factors, it is crucial to examine the coupled fluctuations of urban characteristics and climate, and verify whether their covariation also follows a general scaling function \cite{Hendrick2024}. To this end, we test whether the joint probability density functions $\rho(x,y)$ satisfy the following functional form:
\begin{equation}
    \rho(x , y \thinspace | \langle x \rangle, \thinspace \langle \Delta y \rangle, \sigma_{\Delta y}) = \frac{1} {x^{\lambda}} \thinspace \frac{1}{\sigma_{\Delta y}^{\gamma}} \thinspace  J\Big( \frac{x}{\langle x \rangle^{\delta}} , \frac{y-\langle y_0 \rangle - \langle \Delta y \rangle}{\sigma_{\Delta y}^{\phi}} \Big),
    \label{eq:ansatz2}
\end{equation}
where $J$ is a general scaling function that holds across all cities, and $\phi$, $\gamma$, $\lambda$, and $\delta$ are the scaling exponents.  
As before, we employ equation \eqref{eq:log_mean_std} along with the same clusters identified for the marginal PDFs using the street network ($x{=}n$; $K=3$; Fig.~\ref{fig:2}c for temperature and Fig.~\ref{fig:3}c for $PM$), and rewrite equation \eqref{eq:ansatz2} as a function of urban features: 
\begin{equation}
    \rho(x , y | \langle x \rangle, \langle \ln x \rangle, \sigma_{\ln x}) \thinspace x^{\lambda} \thinspace (\overline{\beta}_k \sigma_{\ln x})^{\gamma} = J\big( x_{rescaled}, y_{rescaled} (x)\big), 
\end{equation}
where $y_{rescaled}(x)$ is defined as in equation \eqref{eq:y_rescaled_kmeans} and the rescaled urban feature is defined as \cite{Hendrick2024}:
\begin{equation}
    x_{rescaled} = \frac{ x }{\langle x \rangle^{\delta}}
    \label{eq:x_rescaled}.
\end{equation}  

After this rescaling, Fig.~\ref{fig:2}f and Fig.~\ref{fig:3}f show that the joint probability distributions of climate with street intersections, $\rho(T,n)$ and $\rho(PM,n)$, achieve a robust collapse onto a single function (visualized in two dimensions as contours of a surface), indicating that once structural differences are accounted for, intra-urban characteristics and climate covariations also follow a common statistical form. 

To identify the scaling exponents that yield the best data collapse, we again employ the residual metric $R$ \cite{Bhattacharjee2001} minimizing $R$ over the two independent scaling exponents $\phi$ and $\delta$, which under the normalization constraint: $\phi=\gamma$ for the climate variable $y$ (see SI Section~\ref{sec:scaling_exponents}), and $\lambda=1$ for the urban feature $x$ (see \cite{Giometto2013}). The insets in Fig.~\ref{fig:2}f and Fig.~\ref{fig:3}f show that the best data collapse occurs at $\phi \simeq \delta \simeq 1$. This result — consistent with the proportionality of moments derivations — implies that the joint distributions preserve their shape across cities and differ only in their characteristic scale — corresponding to the standard deviation for climate variables, and the mean for urban features (e.g., street intersections) — consistently with the results from \cite{Hendrick2024} and our previous findings (see SI Section \ref{sec:scaling_exponents} and SI Fig.~\ref{fig:SI_mean_var}).

\subsection*{Reconciling spatial decay models with observed probability distributions}
\begin{figure}[H]   
    \centering
    \includegraphics[width=\linewidth]{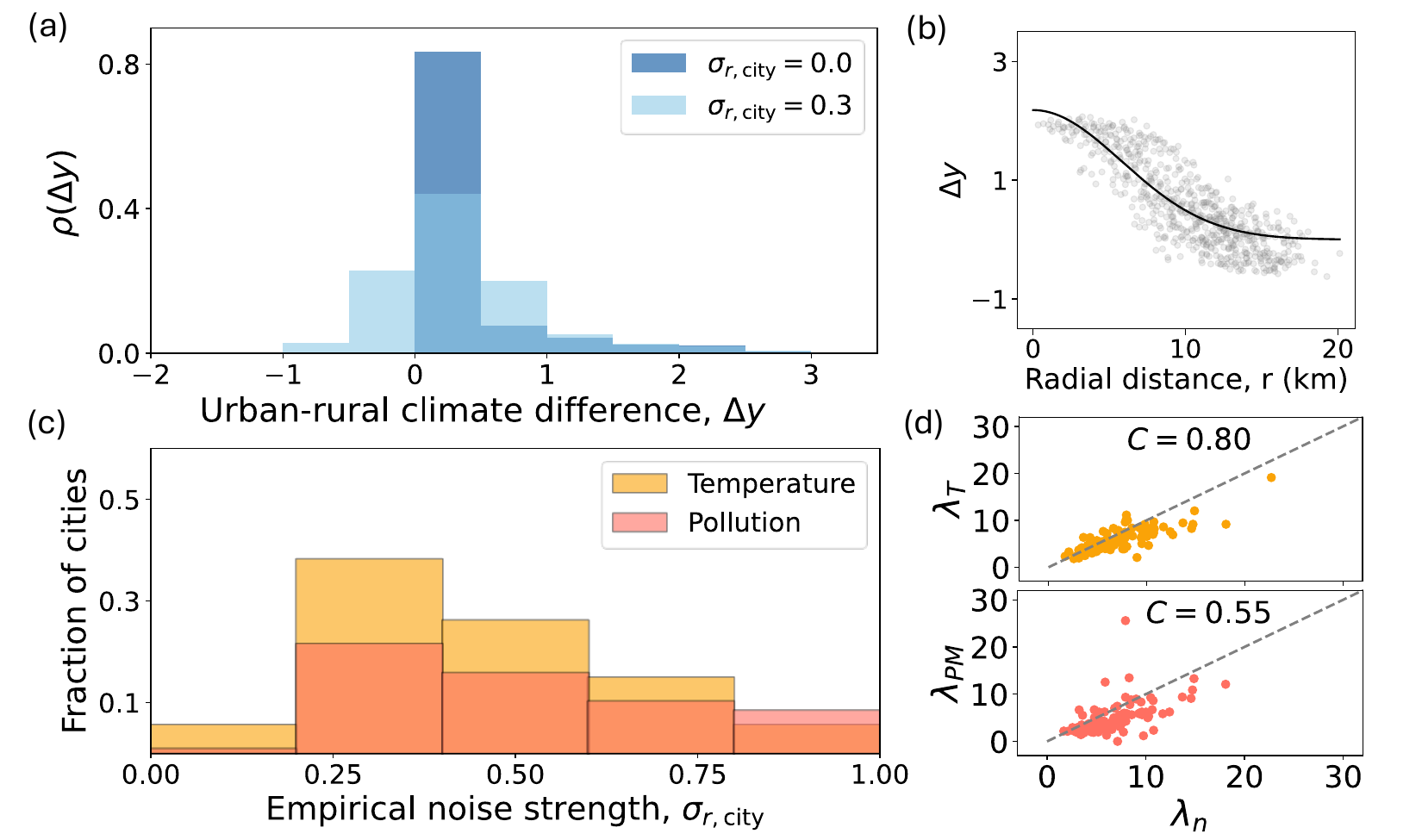} 
    \caption{\footnotesize \textbf{Stochastic radial decay model of climate variables.} (a)  Probability density functions of $\Delta y$ from numerical simulations: when $\sigma_{r,city}=0.0$, the deterministic decay model is recovered, while for $\sigma_{r,city}=0.3$ an approximately Gaussian PDF shape emerges, using a peak value $y_A = 2.5$, decay rate $\lambda_y = 5$~km, and city radius $R = 25$~km. (b) Illustration of the radial decay of climate variables (i.e., urban–rural temperature or pollution differences, $\Delta y$) together with the deterministic decay model fit for Bucharest. (c) Distribution of empirical noise strengths ($\sigma_{r,city}$) of climate variables across all cities. (d) Scatter plots of the street network decay rate ($\lambda_n$) versus both the decay rate of temperature ($\lambda_T$) and particulate matter ($\lambda_{PM}$), along with their corresponding Pearson correlation values $C$, evaluated using two-sided tests, with exact $p$ values of $p = 4.3 \times 10^{-25}$ for temperature and $p = 4.9 \times 10^{-9}$ for particulate matter.}
\label{fig:4}
\end{figure}

Our results reveal that, when appropriately rescaled by urban structure, the probability distributions of climate variables collapse onto an approximately Gaussian form (see Fig.~\ref{fig:2} and Fig.~\ref{fig:3}). Yet, traditional radial decay models — which relate urban-induced microclimate changes to the distance from the city center \cite{Zhou2015,Yu2020} — yield PDFs that deviate from the observed Gaussian behavior (even when polycentric cities are disregarded, see Methods), following instead the form $\rho(y) \propto 1/y$ (see Fig.~\ref{fig:4}a and SI Section \ref{sec:stochastic_decay_model} for the mathematical derivations). Here, we show that this discrepancy arises from the deterministic formulation, which overlooks the substantial fluctuations around the mean decay trend observed in empirical data (see Fig.~\ref{fig:4}b). To show this, we extend the traditional deterministic decay function of urban-rural climate differences by adding an additive stochastic term, that is:
\begin{equation}
\Delta y(r)
= y_{A}\,\exp\!\Bigl(-\tfrac{r^{2}}{2\,\lambda_{y}^{2}}\Bigr)
+ \mathcal{N}\!\bigl(0,{\sigma_{r,\text{city}}}^{2}\bigr),
\label{eq:decay}
\end{equation}
where $r$ denotes the radial distance from the city center (see Methods for definitions), \(y_{A}\) is the climate central value at \(r=0\), \(\lambda_y\) controls the decay rate, and $N\big(0,{\sigma_{r,\text{city}}}^{2}\big)$ represents white noise with standard deviation ${\sigma_{r,\text{city}}}$ — which prompts the PDFs to shift from non-Gaussian to Gaussian shapes (see Fig.~\ref{fig:4}a).
To quantify the strength of the stochastic component ${\sigma_{r,\text{city}}}$ in the data, we compute it as the standard deviation of the residuals — defined as the difference between the observed empirical values and the fitted deterministic decay profile (equation \eqref{eq:decay}) — for each city’s temperature and PM concentration fields (see Methods for details). Figure~\ref{fig:4}c shows the resulting distribution of city-level ${\sigma_{r,\text{city}}}$ values across all cities analyzed.

We further explored whether the decay of urban climate is linked to urban characteristics. To do so, we applied the exponential decay model in equation \eqref{eq:decay} to both street intersections and population counts (see SI Fig.~\ref{fig:SI_decay_results_node}), this time incorporating multiplicative white noise, which reproduces the observed log-normal PDFs of urban features \cite{Hendrick2024} (see SI Fig.~\ref{fig:SI_stochastic_decay_results_node} and SI Section \ref{sec:stochastic_decay_model} for the mathematical derivations). Consistently with our previous results, the street network again emerges as a more reliable descriptor of climate variability than population counts, showing stronger and more significant correlations between decay rates and stochastic noise strengths across cities (see SI Table~\ref{table:SI_correlations}). Notably, cities with stronger street intersection decay rates $\lambda_n$, also tend to exhibit stronger urban climate decay rates for both temperature $\lambda_T$ and pollution $\lambda_{PM}$ (Fig.~\ref{fig:4}d). However, the Pearson correlations between the noise component of street intersections and that of climate variables are moderate for temperature and negligible for pollution (see SI Table~\ref{table:SI_correlations}), suggesting that additional processes (e.g., industrial emissions, atmospheric transport) play an important role.

In summary, these findings provide a conceptual bridge between traditional radial-decay studies and our scaling framework, highlighting the key role of urban structure in shaping intra-urban climate variability.

\section*{Discussion}\label{sec:discussions}
We have shown that marginal and joint probability distributions of intra-urban air temperature and PM concentrations across 142 cities, when rescaled by characteristic scales derived from the urban structure, collapse into general scaling functions that approximate a Gaussian form (see Figs.~\ref{fig:2}--\ref{fig:3}). While inspired by finite-size scaling ideas from statistical physics \cite{fisher1972scaling,cardy2012finite}, the observed collapse reflects a location–scale statistical regularity of intra-urban climate fields which deviates from the traditional FSS ansatz, with size-dependence replaced by the variance of the climatic variable (equation \eqref{eq:ansatz}). As such, this study advances existing scaling approaches based on city‐level metrics \cite{Li2020,Sobstyl2018,Zhao2014,Manoli2019,Han2014,Lamsal2013}, by extending them to the entire intra‐urban variability of climate, and further shows that climatic variables and their underlying urban structures follow different statistical forms \cite{Hendrick2024}. As a result, we overcome long-standing problems related to the definition of city boundaries \cite{Cottineau2017, Arcaute2015} and the neglect of local urban heterogeneities \cite[e.g.,][]{Volpati2018}, which would otherwise mask important microclimatic effects \cite{Vardoulakis2011,Qiu2017}. This framework does not aim to model specific physical or chemical processes but rather to uncover generic spatial statistical regularities that emerge across cities with diverse morphologies, climatic conditions, and cultural contexts. Similarly, recent work has shown that urban–rural differences in the temporal occurrence of heat events across cities can be predicted using a standardized Gaussian framework that combines the mean UHI with rural temperature variance and persistence \cite{Liao2025}. This parallel suggests that both temporal and spatial urban climate fluctuations can be described within a common statistical paradigm.

While the general scaling holds across all cities, clustering reveals inter-city differences beyond urban morphology — such as regional climate, local emissions regulations or topography (see SI Table~\ref{table:SI_cluster_alpha_beta_absbeta} and SI Section~\ref{sec:clustering_effect} for details). In particular, rescaling the climate fields based on street network information led to $K=3$ clusters (Figs.~\ref{fig:2}c--\ref{fig:3}c) that improve the data collapse but do not change the underlying shape function $G$, which is shared across all cities.

The methodological robustness of our approach is confirmed by several sensitivity tests. Specifically, we demonstrate scale independence by recalculating our results using square grid cells of different sizes (see SI Fig.~\ref{fig:SI_scales} and Methods), and confirm the general validity by repeating the analysis with independent datasets for both temperature and pollution (SI Figs.~\ref{fig:SI_double_check_T}- \ref{fig:SI_double_check_PM} and SI Section~\ref{sec:alternative_datasets} for details).

We further show that traditional radial decay models for urban climate \cite{Zhou2015,Yu2020} produce heavy-tailed probability density functions that deviate from the near-Gaussian behavior observed here. By adding an additive stochastic term to the radial decay model — empirically derived and quantified from the data — we are able to reproduce the observed distributions. This reconciliation suggests that intra-urban climate is shaped by the interplay between the predictable decay with distance of infrastructure density — explained, for example, by gravitational models \cite[e.g.,][]{li2021singularity} — and the randomness inherent in heterogeneous urban morphologies.

In our study, we used two simple urban predictors — street intersections and population counts — because they are globally available, consistently defined, and reflect built form and human activities, which are key drivers of urban heat and pollution. More detailed covariates (e.g., vegetation, land cover) may improve local predictability where available, but would reduce the generality of global cross-city scaling analyses \cite{Barthelemy2024}. Notably, we find that the street network is a more reliable predictor of intra-urban climate variability than population counts. This is reflected in the more consistent fitted parameters ($\alpha$, $\beta$) in equation \eqref{eq:log} (SI Fig.~\ref{fig:SI_log_results}), the tighter data collapse achieved with fewer clusters $K$ (SI Fig.~\ref{fig:SI_clustering_kmeans_popu_vs_node}), and the stronger correlations between decay rates and stochastic noise strengths across cities (SI Table~\ref{table:SI_correlations}). From a physical perspective, this is explained by the fact that urban climates depend on the geometry and energy budgets of different urban surfaces, which are directly linked to infrastructure characteristics — while population is only an indirect measure of urban form.

Our results are consistent with the findings by La Porta et al. \cite{LaPorta2024}, who demonstrated that the probability distributions of total $CO_2$ and $PM$ across different population ranges share a common shape, once normalized by each city’s total population. Thus, while their work reveals general fluctuations at the city scale, our results extend their insight to the intra-urban level, uncovering common behaviors driven by heterogeneous spatial structures, within and across cities. In the context of intra-city variations, Shreevastava et al. \cite{Shreevastava2019a,Shreevastava2019b} investigated the patterns of land surface temperature and identified urban heat islets — i.e., clusters of high surface temperatures — characterized by power-law size distributions. Our results complement their findings by demonstrating that the entire variability of urban climate characteristics — defined across uniform spatial units rather than heat islets — exhibits general scaling patterns.

This work has several limitations that open up clear next steps. First, we used annual means to ensure comparability across cities worldwide and to emphasize persistent structural differences in intra-urban climate, while minimizing the influence of transient events such as heatwaves episodes. However, annual-mean temperature and PM fields overlook temporal dynamics and further research could incorporate changes on diurnal \cite{Manning2018, Zhou2016} and seasonal \cite{Zhang2015, Manoli2020} timescales, thereby enabling a full characterization of the corresponding PDFs over time.  Moreover, PM concentrations show a slightly weaker collapse than temperature, which is consistent with the fact that air pollution is influenced by advection and localized emission sources - factors that go beyond urban morphology. A key challenge is therefore to include richer descriptors without losing the generality and transferability of our results.

In conclusion, this study provides a statistical framework, inspired by FSS theory, that fully describes intra-city fluctuations of urban microclimate and links them to simple measures of urban form. The simplicity of the approach makes it particularly valuable for cities with limited computational capacity, as well as for incorporating climate information into reduced-complexity models of urban systems, from transport to economic and energy systems \cite{barthelemy2019statistical}. This opens up the opportunity to (i) benchmark urban climate models by testing their ability to reproduce the observed scaling of marginal and joint PDFs; (ii) downscale regional simulations to the intra-urban level; and (iii) obtain first-order estimates of spatial patterns of urban climate in cities where detailed monitoring/modeling is unavailable. Our framework further advances the theoretical understanding of how urban form shapes intra-urban climate dynamics and establishes a direct bridge between urban complexity theories and the quantitative description of urban climate variability.
Given the urgency of climate mitigation and adaptation, our results suggest that cities worldwide should speed up the uptake of zero-emission technologies as urban structure, mobility, and pollution are inherently coupled. Also, considering the observed covariation of population with the intensity of urban climate perturbations, there is a pressing need for strategies that balance the positive effects of higher urban densities (fostering, for example, active mobility \cite{Cerin2022}) with the increased exposure of urban residents to climate-related risks \cite{Huang2023}.

\section*{Methods}
\textbf{Data Sources.} This study includes 142 cities for which urban climate simulations \cite{temperature_urbclim} were available. Although the sample mostly contains European cities — reflecting the coverage of the urban climate database — it spans all major climate zones and levels of urban development across the world, providing a diverse global sample. The selection was constrained solely by the availability of simulated air temperature data, without applying additional filters. All cities included had sufficient spatial extent to ensure meaningful intra-urban comparisons.

We used mean daily ambient temperature estimates at 2~m above ground with a 100 × 100 m spatial resolution produced by the UrbClim model \cite{temperature_urbclim} via the VITO UrbClim FTP server (see Data and Code Availability).
UrbClim was developed by the Flemish Institute for Technological Research and it integrates a land surface scheme with a three-dimensional atmospheric boundary layer module, incorporating land cover, soil sealing, vegetation, and meteorological data from the fifth-generation European Centre for Medium-Range Weather Forecasts (ECMWF) Reanalysis. For this study, annual averages of temperature estimates were computed for the entire year 2017.

The annual average $PM_{2.5}$ concentrations are derived from a machine-learning-based global dataset \cite{pm25data_global}. The dataset provides daily $PM_{2.5}$ estimates at a 1 km resolution for the period 2017–2022, generated using a model that integrates multiple data sources, including satellite aerosol optical depth (AOD) retrievals (e.g., MODIS MAIAC and GEOS-FP), ground-based air quality monitoring stations (9500 globally), meteorological variables (e.g., temperature, humidity, wind speed), emissions inventories, and land use information. A spatiotemporally enhanced machine learning model (4D-STET) is used to fill gaps in AOD observations and to account for complex relationships between pollution, meteorology, and surface features. For this study, we use the annual average for 2022 to characterize particulate matter exposure across cities.

The street networks are extracted from OpenStreetMap (OSM) \cite{OpenStreetMap}, an open-access, collaborative mapping project that provides continuously updated, editable maps of the world. The data is processed using OSMnx \cite{Boeing2017} to analyze the structural variability of urban transport networks, offering detailed information on streets, intersections, buildings, and other infrastructure. In this study, we specifically use this dataset to analyze street network intersections. While OSM provides the most comprehensive global database of urban street networks, its completeness can vary across cities — particularly in historical cores. However, our analysis focuses on statistical properties aggregated over thousands of grid cells per city, which mitigates the impact of local incompleteness.

Finally, we use global population count data from the WorldPop dataset \cite{population_data}, which integrates remote sensing, census data, and geospatial technologies to produce high-resolution population distribution maps. These datasets provide consistent and comparable population estimates across urban and rural areas, with resolutions as fine as 100 x 100 meters in some regions. For this study, we use the unconstrained global mosaic population dataset at 1 km resolution for the year 2020, which was created using a top-down disaggregation approach based on statistical models and satellite-derived covariates. The dataset is designed to provide consistent 1 km resolution global population estimates, making it suitable for large-scale comparative urban studies.

To validate the robustness of our findings, we compare the results with alternative datasets. For $PM_{2.5}$ concentration we use a dataset covering Europe downloaded from the European Environmental Agency (EEA) for the year 2022 \cite{pm25data_europe}, and for temperature we use a dataset of global land surface temperature (LST) for the year 2013 \cite{temperature_observational} (see SI Section \ref{sec:alternative_datasets} for details).

\textbf{Data Integration.} To ensure consistent spatial resolution across cities, we aggregate all data into a regular square grid. Each square cell has a side length size $l = 1000m$ (Fig.~\ref{fig:1}b), partitioning the urban area into spatially uniform units across all cities for the analysis, thereby defining urban features as density measures. Within each cell, we compute the mean of all temperature and $PM_{2.5}$ measurements, and the sum of all street intersections and population counts. A sensitivity analysis with different side length sizes $l$ is provided in the SI (see SI Fig.~\ref{fig:SI_scales}). The cells of the different scales have been delimited to keep spatial coherence: four 500~m cells form one of the 1~km cells used in the previous figures, and four 1~km cells aggregate to form a 2~km cell. Note that population and $PM_{2.5}$ data are available only at 1~km resolution; therefore, the 500~m were performed only for temperature and street-network data.

We define a distinction between urban and rural areas using the street network. Specifically we consider as urban grid cells where there is at least one intersection ($n\geq1$), and rural where there is only one intersection ($n=1$) to avoid water zones. The average rural climate value $\langle y_0 \rangle$ is computed at $1000~m$ resolution — where the agreement with the UrbClim classification of rural areas \cite{temperature_urbclim}, based on land use and built-environment characteristics, is highest. This urban–rural distinction is used to remove the average rural climate signal and to isolate the mean and scale of urban climate in a consistent manner across all spatial resolutions $l$ for the ansatz (equation \eqref{eq:ansatz}), which reflects the city’s perturbation of the background climate field. This procedure ensures that the rural climate baseline remains physically consistent across scales, while allowing the urban variability to be assessed at multiple spatial resolutions.

We adopt the spatial domain provided by our UrbClim temperature dataset \cite{temperature_urbclim} for all cities. While previous research suggests that city structure follows consistent radial profiles that scale with population size \cite{Remi2020}, allowing for rescaled urban domains based on population-dependent functions, we find that applying such rescaling does not alter our results. In fact, we tested multiple variants of the rescaled radial extent and observed nearly identical data collapses, as long as the same criterion was applied consistently across all cities. 


\textbf{Statistical Analysis.} We employ kernel density estimation (KDE) \cite{Silverman2018} because it provides a flexible and non-parametric approach to estimate and derive probability density functions (PDFs) for urban climate and urban characteristics variables. PDFs are computed for population counts, street network intersections, air temperature and PM concentrations, providing insights into the spatial distribution and variability of these variables within cities. In addition, the quality of the data collapse — both for marginal and joint probability density functions — is quantitatively assessed using the residual-based metric proposed in \cite{Bhattacharjee2001}, which evaluates the overall distance between rescaled curves.


\textbf{Clustering.} We employ the K-means clustering method \cite{Macqueen1967} because it provides a simple and effective way to cluster cities based on similarity in their data collapse response, while requiring no prior assumptions about cluster shape or distribution. For each city, we first computed the rescaled climate distribution using a global intercept $\overline{\alpha}$ and slope $\overline{\beta}$ values averaged across all 142 cities ($K=1$, see SI Fig.~\ref{fig:SI_collapse_clustering}) — and summarize its form by the coordinates of its maximum density point, $(x_{peak},y_{peak})$. Thus, each city is represented by a two-component feature vector. We run K-means for  $k=1,\dots,10$, which partitions cities by minimizing the within-cluster sum of squared Euclidean distances in the two-dimensional \((x_{\rm peak},y_{\rm peak})\) space. The optimal number of clusters is selected using the elbow of the within-cluster sum-of-squares, expressed as explained variance. Using the street network led to $K=3$ for both temperature and pollution (see SI Fig.~\ref{fig:SI_clustering_kmeans_popu_vs_node}). Finally, to ensure the tightest possible collapse of the rescaled curves, we recomputed the Bhattacharjee \emph{et al.}\ residual metric $R$ \cite{Bhattacharjee2001} and found that only 4.2\% for temperature (6 cities) and 9.9\% for pollution (14 cities) changed cluster assignment, confirming the robustness of our grouping and very low computational cost.

\textbf{Radial Decay Model.} We defined city center as the geographic centroid of all the street intersection coordinates, computed as the mean of their latitude and longitude values (independently of the regular square grid system). For each grid cell centroid (longitude, latitude) within the gridding system, the radial distance $r$ was calculated as the haversine (i.e., geodesic) distance between its centroid coordinates and the city center ($r=0$). This geodesic formulation ensures that distances are physically consistent across cities located at different latitudes and longitudes, providing a reproducible reference framework for the radial decay analysis.
To ensure the validity of the radial decay framework — which assumes approximate circular symmetry — we excluded cities with poor fits for the street intersection decay model ($R^2<0.1$), keeping 107 out of 142 cities (76\%) for this spatial decay analysis. An additional 12 cities were excluded from the particulate matter ($PM$) analyses due to implausible decay rates.

To test the white-noise assumption in the stochastic climate radial decay model, we divided each city into concentric radial belts, each containing the same number of data points. We then calculated the standard deviation of the decay-model residuals ($\sigma_r$) within each belt. This procedure allowed us to test whether residual fluctuations were approximately uniform across distance (see SI Fig.~\ref{fig:SI_illustration_decay_cities_noise}). SI Fig.~\ref{fig:SI_decay_results} and SI Fig.~\ref{fig:SI_decay_results_node} provide fit diagnostics — residual histograms and $R^2$ statistics — for both climate variables and urban features.

Note that the selection of a subset of cities for the decay model analysis, does not alter the generality of the scaling functions describing the marginal and joint PDFs — which are the main focus in our study.

\section*{Data Availability}
The datasets utilized in this study are publicly available from the following sources: population count data from \url{https://hub.worldpop.org/}, street network data from \url{https://www.openstreetmap.org/}, temperature data from \url{https://provide.marvin.vito.be/ftp/compressed_daily/} and \url{https://www.earthdata.nasa.gov}, and PM2.5 data from \url{https://doi.org/10.5281/zenodo.10800980} and \url{https://www.eea.europa.eu/en}.

\section*{Code Availability}
The Jupyter notebooks and associated scripts necessary to reproduce the results of this paper are accessible on GitHub at \url{https://github.com/urbes-team/scaling_intra_urban_climate}.

\section*{Acknowledgements}
The authors would like to thank Andrea Rinaldo for the inspiring conversations at the beginning of this research.

\section*{Author Contributions}
G.M. conceived and supervised the project. M.D.S. devised and performed the analyses and drafted the manuscript with inputs from M.H. and G.M. All authors interpreted the results, provided feedback that helped shape the analysis, and contributed to writing the manuscript.


\section*{Competing Interests}
The authors declare no competing interests.

\section*{Figures and Tables}

\begin{figure}[H]   
    \centering
    \includegraphics[width=\linewidth]{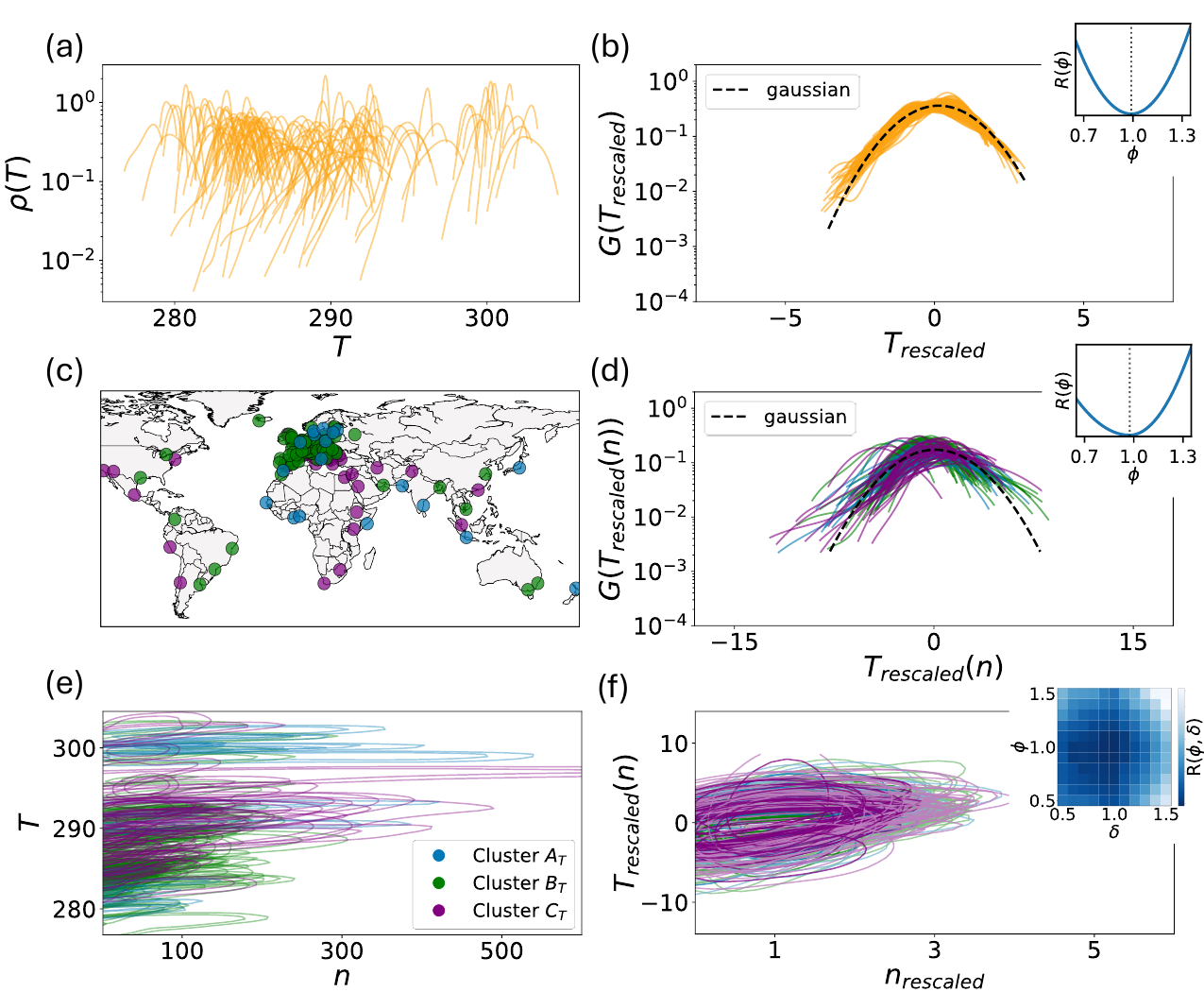} 
    \caption{\footnotesize 
    \textbf{Scaling intra-urban temperature ($T$) variations and covariations with street network intersections ($n$).}
    (a) Distinct PDFs of temperature observed across the analyzed cities, where each curve represents one city.
    (b) After rescaling using the empirical climate data according to equations \eqref{eq:ansatz}-\eqref{eq:y_rescaled}, these PDFs collapse onto a common scaling function $G$. A Gaussian fit (dashed black line) indicates that $G$ is well approximated by a normal distribution. The inset shows the residual $R(\phi)$.
    (c) Global distribution of cities and their grouping into three clusters. 
    (d) Collapse of the marginal PDFs after rescaling using the street network information according to equations \eqref{eq:ansatz_rescaled_kmeans}-\eqref{eq:y_rescaled_kmeans}. As for panel b, the Gaussian fit (dashed black line) and residual $R(\phi)$ (inset) are also shown. 
    (e) Joint PDFs for temperature and street network intersections for all cities analyzed. Colors refer to the different clusters illustrated in panel c.
    (f) After both variables are rescaled using equation \eqref{eq:ansatz2}, the data collapse follows a common joint scaling function. The inset shows the residual $R(\phi,\delta)$.}
\label{fig:2}
\end{figure}

\begin{figure}[H]   
    \centering
    \includegraphics[width=\linewidth]{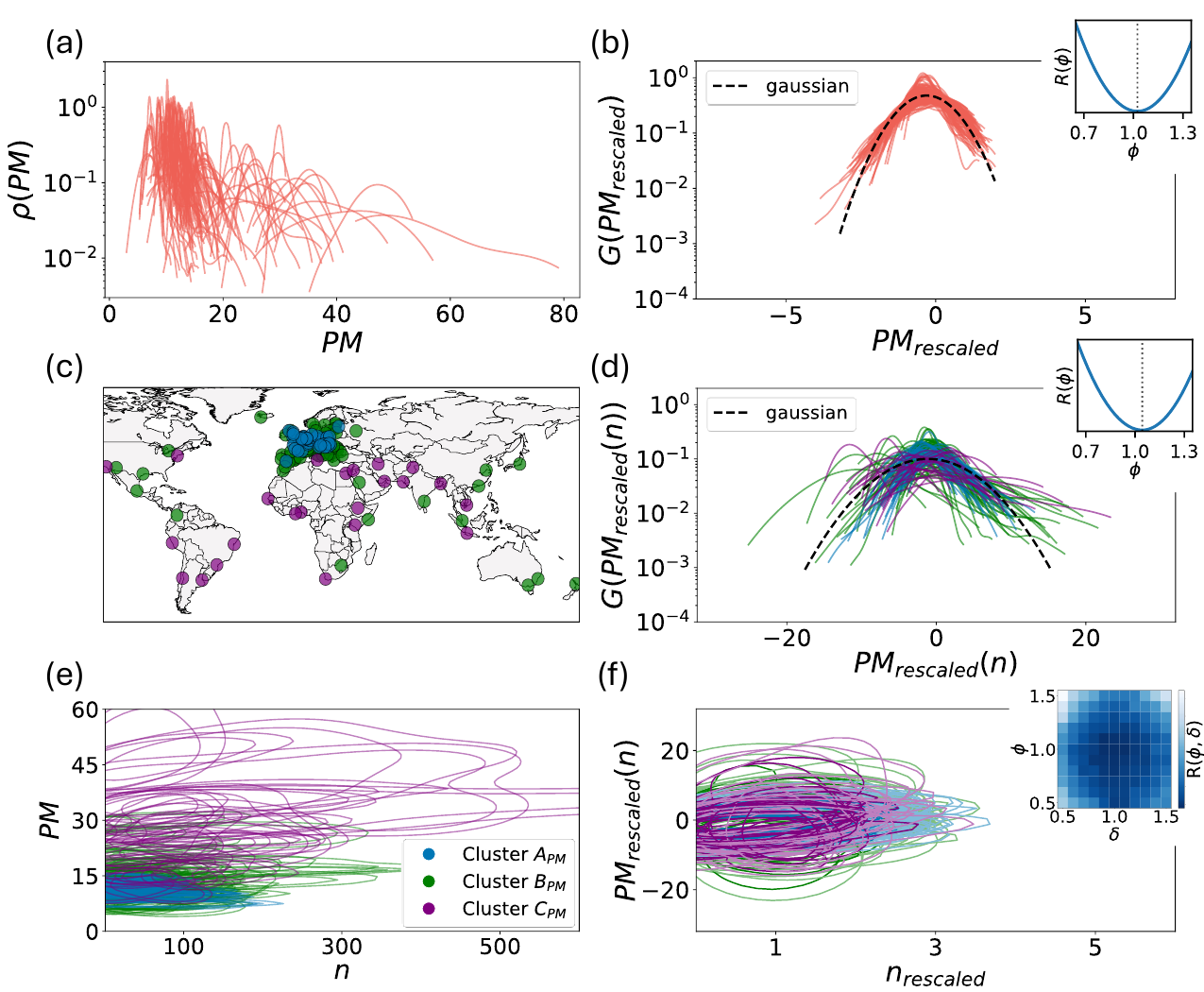} 
    \caption{\footnotesize 
    \textbf{Scaling intra-urban air quality ($PM$) variations and covariations with street network intersections ($n$).}
    (a) Distinct PDFs of particulate matter concentrations observed across the analyzed cities, where each curve represents one city.
    (b) After rescaling using the empirical climate data according to equations \eqref{eq:ansatz}-\eqref{eq:y_rescaled}, these PDFs collapse onto a common scaling function $G$. A Gaussian fit (dashed black line) indicates that $G$ is well approximated by a normal distribution. The inset shows the residual $R(\phi)$.
    (c) Global distribution of cities and their grouping into three clusters. 
    (d) Collapse of the marginal PDFs after rescaling using the street network information according to equations \eqref{eq:ansatz_rescaled_kmeans}-\eqref{eq:y_rescaled_kmeans}. As for panel b, the Gaussian fit (dashed black line) and residual $R(\phi)$ (inset) are also shown.
    (e) Joint PDFs for temperature and street network intersections for all cities analyzed. Colors refer to the different clusters illustrated in panel c.
    (f) After both variables are rescaled using equation \eqref{eq:ansatz2}, the data collapse follows a common joint scaling function. The inset shows the residual $R(\phi,\delta)$.}
\label{fig:3}
\end{figure}

\bibliographystyle{unsrtnat}  
\bibliography{bib}            

\newpage
\section*{Supplementary Information to: \\ ``Scaling intra-urban climate fluctuations''}

{\large Marc Duran-Sala$^1$, Martin Hendrick$^1$, Gabriele Manoli$^{1,*}$} \\
$^{1}$Laboratory of Urban and Environmental Systems, \'{E}cole Polytechnique F\'{e}d\'{e}rale de Lausanne, Lausanne, Switzerland \\
$^{*}$Corresponding author: gabriele.manoli@epfl.ch






\newcommand{\beginsupplement}{%
        \setcounter{table}{0}
        \renewcommand{\thetable}{S.\arabic{table}}%
        \setcounter{figure}{0}
        \renewcommand{\thefigure}{S.\arabic{figure}}%
        \setcounter{equation}{0}
        \renewcommand{\theequation}{S.\arabic{equation}}%
        \setcounter{section}{0}
        \renewcommand{\thesection}{S.\arabic{section}}%
     }

\beginsupplement


\section{Scaling exponents}
\label{sec:scaling_exponents}
This derivation follows the same approach used in scaling analyses, where normalization and moment proportionality requirements impose constraints on scaling exponents  and provide a physical interpretation of the characteristic scale (e.g., see \cite{Giometto2013}, although applied to a different scaling form). 

\vspace{1em}

\noindent
\textbf{Normalization.} Here, we show that the normalization requirement $\int_{-\infty}^{\infty} \rho(y)\, dy = 1$ for climate distributions of the form  
\begin{equation}
\rho(y \,|\, \langle \Delta y \rangle, \sigma_{\Delta y}) = \frac{1}{\sigma_{\Delta y}^{\gamma}} \, G \left( \frac{(y - \langle y_0\rangle) - \langle \Delta y \rangle}{\sigma_{\Delta y}^{\phi}} \right)
\label{eq:ansatz_SI}
\end{equation}
constraints the possible values of the scaling exponents \(\gamma\) and \(\phi\).

By performing the change of variables  
$$
u = \frac{y - \langle y_0\rangle - \langle \Delta y \rangle}{\sigma_{\Delta y}^{\phi}},
\quad \text{so that} \quad dy = \sigma_{\Delta y}^{\phi} \, du,
$$
the normalization condition becomes:  

\begin{equation}
\int_{-\infty}^{\infty}\rho(y\,|\,\langle \Delta y \rangle, \sigma_{\Delta y})\,dy 
= \sigma_{\Delta y}^{\phi-\gamma} \int_{-\infty}^{\infty} G(u)\,du = 1 \Rightarrow \phi=\gamma
\end{equation}

\vspace{1em}

\begin{figure}[H]   
    \centering
    \includegraphics[width=\linewidth]{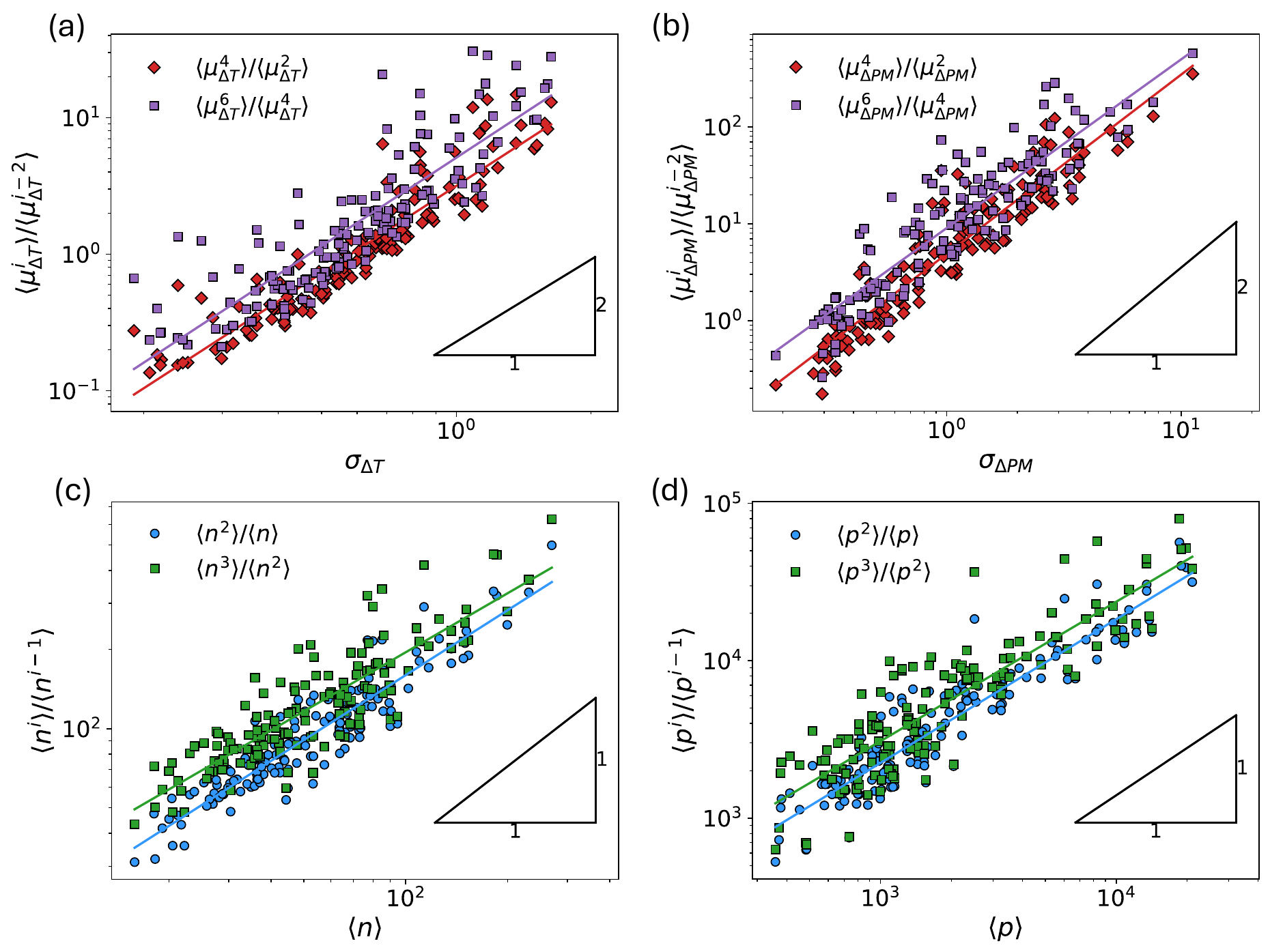} 
    \caption{\footnotesize Independent test that further corroborates that intra-urban climate distributions are fully characterized by their mean and variance (a-b), and intra-urban features distributions are fully characterized by their mean (c-d). Panels (a) and (b) show the ratios of successive even moments versus the standard deviation in log-log scale for temperature and pollution, respectively, with slopes close to two - further supporting the ansatz and the data collapse results. Here, $\mu_{\Delta y} = (y-\langle y_0 \rangle - \langle \Delta y \rangle)$. Panels (c) and (d) show the ratios of successive moments versus the mean in log-log scale street intersections and population counts, respectively, with slopes close to one - further supporting the ansatz and the data collapse results in \cite{Hendrick2024}.}
\label{fig:SI_mean_var}
\end{figure}

\noindent
\textbf{Proportionality of moments.} Let us look at a generic \(j\)-th central moment, defined for convenience as:
\begin{equation}
\langle \bigl[(y - \langle y_0\rangle) - \langle \Delta y \rangle \bigr]^j \rangle
= \int_{-\infty}^{\infty} \bigl[(y - \langle y_0\rangle) - \langle \Delta y \rangle \bigr]^j \,\rho(y)\,dy.
\label{eq:central_moment}
\end{equation}
Using the same change of variable $u$ as before, one gets:

\begin{align}
\langle \bigl[(y - \langle y_0\rangle) - \langle \Delta y \rangle\bigr]^j \rangle
&= \int_{-\infty}^{\infty} \bigl(\sigma_{\Delta y}^\phi\,u\bigr)^j
\frac{1}{\sigma_{\Delta x}^\gamma} G(u)\,\sigma_{\Delta y}^\phi\,du
\nonumber\\[6pt]
&= \sigma_{\Delta y}^{\phi j + \phi - \gamma}
\int_{-\infty}^{\infty} u^j\,G(u)\,du,
\end{align}
therefore, as we previously showed that \(\gamma = \phi\), we can write:
\begin{equation}
\langle \bigl[(y - \langle y_0\rangle) - \langle \Delta y \rangle\bigr]^j \rangle
= \sigma_{\Delta y}^{\phi j}\,\int_{-\infty}^{\infty} u^j\,G(u)\,du = K_j\,\sigma_{\Delta y}^{\phi j},
\label{eq:moments_scaling}
\end{equation}

where $K_j$ are dimensionless constants that depend only on the common shape $G$. More explicitly:
\begin{itemize}
\item For \(j=1\), we get the first central moment
\(\langle (y - \langle y_0\rangle - \langle \Delta y \rangle)\rangle = \sigma_{\Delta y}^{\phi}\int u\,G(u)\,du.\) By requiring \(\int u\,G(u)\,du=0\), we ensure that \(\langle \Delta x \rangle\) is the mean shift.
\item For \(j=2\), we get 
\(\langle (y - \langle y_0\rangle - \langle \Delta y \rangle)^2\rangle = \sigma_{\Delta y}^{2\phi}\int u^2\,G(u)\,du.\) By requiring \(\int u^2\,G(u)\,du=1\) and \(\phi=1\), we identify \(\sigma_{\Delta x}\) as the standard deviation.
\end{itemize}
From equation \eqref{eq:moments_scaling}, we get
\[
\frac{\left\langle(\Delta y-\langle\Delta y\rangle)^{j}\right\rangle}
{\left\langle(\Delta y-\langle\Delta y\rangle)^{j-2}\right\rangle}
= \frac{K_j}{K_{j-2}}\;\sigma_{\Delta y}^{2\phi}\quad(j\ge 4).
\]
Thus, the proportionality of successive even moments ratios to $\sigma_{\Delta y}$ provides an independent test. On log–log axes, the expected slopes are $2\phi$. Since we empirically found on our data slopes near two (see Fig.~\ref{fig:SI_mean_var}), this implies that $\phi\simeq 1$ for both temperature and pollution, meaning $(\langle \Delta y\rangle,\sigma_{\Delta y})$ fully characterize the distributions across cities, consistent and further supporting the ansatz in equation \eqref{eq:ansatz} and the data collapse results from Fig.~\ref{fig:2} and \ref{fig:3}. We focus on the proportionality of successive even moments ratios since $G$ is approximately symmetric and all odd $K_{2k+1}$ are nearly zero. Thus, the proportionality of successive moments ratios (i.e., $\langle(\Delta y-\langle\Delta y\rangle)^{j}\rangle/\langle(\Delta y-\langle\Delta y\rangle)^{j-1}\rangle$) are numerically noisy.

Similarly, for urban features the successive moment ratios scale as $\langle x^j \rangle / \langle x^{j-1} \rangle \propto \langle x \rangle^{\delta}$. Thus, the proportionality of successive moment ratios to $\langle x \rangle$ with slopes near one on log-log scale (see Fig.~\ref{fig:SI_mean_var}) implies that $\delta\simeq 1$, providing an analogous independent test that further corroborates the ansatz and the data collapse results from \cite{Hendrick2024}.


\section{Derivation of the general urban climate scaling function}
\label{sec:urban_climate_universal_scaling_form}

Previous study~\cite{Hendrick2024} has shown that urban variables \( x \) (such as population, street intersections, \emph{etc.}) can be described through their probability density function \( \rho(x) \) according to the scaling form:
\begin{equation}
  \rho(x \mid \langle x \rangle) = x^{-\lambda} \, F \left( \frac{x}{\langle x \rangle^\delta} \right),
  \label{eq:fss_y}
\end{equation}
with scaling exponents \( \lambda, \delta \) and mean \( \langle x \rangle \).

For simplicity, we can reasonably assume that the urban feature \( x \) follows a lognormal distribution \cite{Hendrick2024}:

\begin{equation}
  \ln x \sim \mathcal N\!\bigl(\langle \ln x \rangle,\;\sigma_{\ln x}^{2}\bigr)
  \;\;\Longrightarrow\;\;
  \rho(x)
   = \frac{1}{x\sqrt{2\pi\sigma_{\ln x}^{2}}}\,
     \exp\!\!\left[-\frac{(\ln x-\langle \ln x \rangle)^{2}}{2\sigma_{\ln x}^{2}}\right].
  \label{eq:lognormal_y}
\end{equation}
In this paper we use a logarithmic relationship to link intra-urban climate variability $\Delta y$ with an urban feature $x$ as:
\begin{equation}
  \Delta y = \alpha + \beta \ln x\;
  \quad\Longleftrightarrow\quad
  x = \exp\!\Bigl[\tfrac{\Delta y-\alpha}{\beta}\Bigr],
  \label{eq:loglinear}
\end{equation}
where $\alpha$ and $\beta$ take cluster‑specific averages $(\bar\alpha_k,\bar\beta_k)$. From \eqref{eq:loglinear} we get:
\[
\left|\frac{dx}{d\Delta y}\right|
   = \frac{1}{\beta}\exp\!\Bigl[\tfrac{\Delta y-\alpha}{\beta}\Bigr]
   = \frac{x}{\beta}.
\]
The PDF of $\Delta y$ is therefore:
\begin{align}
\rho(\Delta y)
 &= \rho(x)\;\left|\frac{dx}{d\Delta y}\right| \nonumber\\
 &= \frac{1}{\beta\sqrt{2\pi\sigma_{\ln x}^{2}}}\,
    \exp\!\Bigl[-\tfrac{(\ln x-\langle \ln x \rangle)^{2}}{2\sigma_{\ln x}^{2}}\Bigr]
    \Bigg|_{\,\ln x=\frac{\Delta y-\alpha}{\beta}} \nonumber\\
 &= \frac{1}{\beta\sqrt{2\pi\sigma_{\ln x}^{2}}}\,
    \exp\!\Bigl[-\tfrac{\bigl(\frac{\Delta y-\alpha}{\beta}-\langle \ln x \rangle\bigr)^{2}}
                        {2\sigma_{\ln x}^{2}}\Bigr].
 \label{eq:pdf_dx_raw}
\end{align}
Using the mean $\langle\Delta y\rangle = \alpha + \beta \langle lnx \rangle$ and standard deviation $\sigma_{\Delta y} = \beta\sigma_{\ln x}$, \eqref{eq:pdf_dx_raw} simplifies to:
\begin{equation}
  \rho(\Delta y)
  = \frac{1}{\sigma_{\Delta y}\sqrt{2\pi}}\,
    \exp\!\Bigl[-\tfrac{(\Delta y-\langle\Delta y\rangle)^{2}}
                       {2\sigma_{\Delta y}^{2}}\Bigr].
  \label{eq:pdf_dx_gaussian}
\end{equation}
Thus, we shown that $\Delta y$ is normally distributed.

By defining the location and scale parameters as:
\begin{equation}
\langle\Delta y\rangle = \alpha+\beta \langle \ln x \rangle \quad \text{and} \quad S := \sigma_{\Delta y} = \beta\,\sigma_{\ln x},
\end{equation}
that are both computed from the urban feature, we can rewrite~\eqref{eq:pdf_dx_gaussian} as:
\begin{equation}
  \rho(\Delta y)
  = S^{-1}\;
    G\!\Bigl(
        \tfrac{\Delta y-\langle\Delta y\rangle}{S}
      \Bigr),
  \label{eq:fss_form}
\end{equation}
where the general scaling function
\[
G(u)=\frac{1}{\sqrt{2\pi}}e^{-u^{2}/2},
\]
is the same for every city once the data are rescaled by $\langle\Delta y\rangle$ and ~\(S\).

Equation \eqref{eq:fss_form} corresponds exactly to the scaling form given in equation \eqref{eq:ansatz} we used in the main text and that constitutes our key expression to emphasize the general form across cities. We deliberately avoid explicitly referring to the Gaussian distribution in the main text, and instead present a more general formulation through the function $G$. This choice reflects that a perfect Gaussian shape would only be expected in the infinite-size limit, according to the central limit theorem, and under the assumption that the random variables are strictly independent and identically distributed (i.i.d.).




        
        


\section{Clustering analysis}
\label{sec:clustering_effect}
\begin{figure}[H]   
    \centering
    \includegraphics[width=\linewidth]{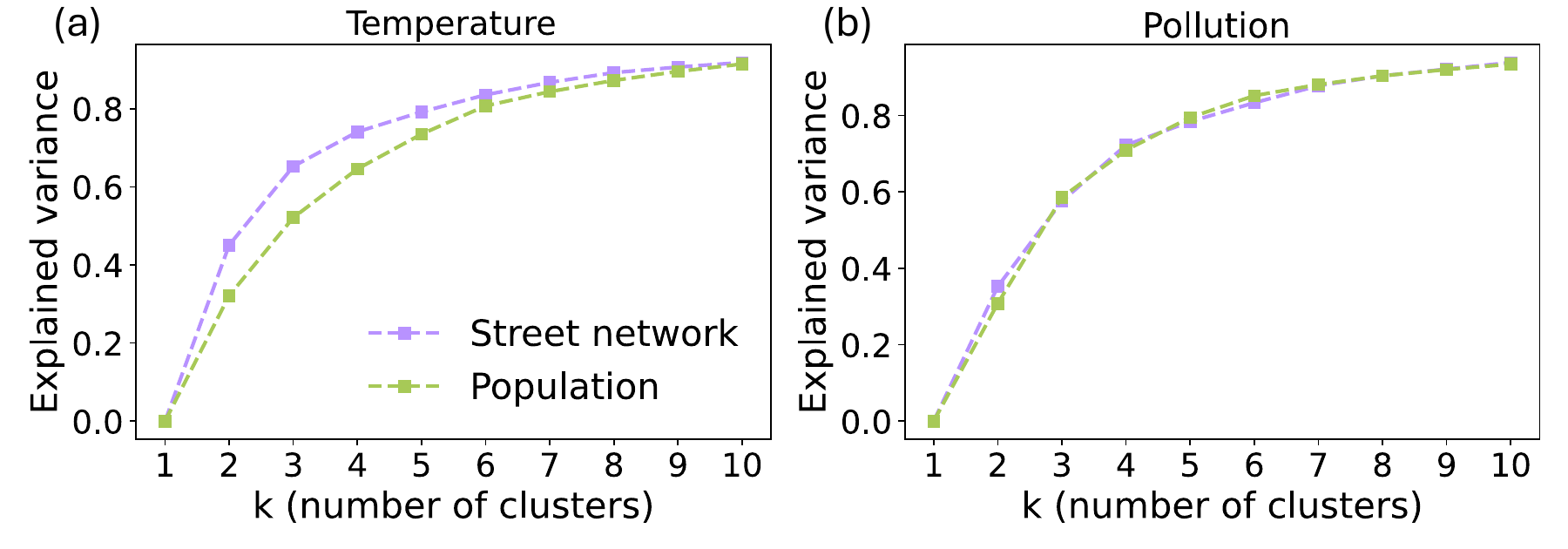} 
    \caption{\footnotesize Explained variance obtained from $k$–means clustering as functions of the number of clusters $k$ using street network and population data for (a) temperature and (b) pollution. At each $k$, the street network clustering explains a higher fraction of explained variance than population for temperature, and a comparable fraction for pollution. In both cases, the variance explained by street network clustering increases sharply up to $k \approx 3$ - reaching about $\sim60\%$ - and then plateaus, indicating diminishing returns from adding more clusters.}
\label{fig:SI_clustering_kmeans_popu_vs_node}
\end{figure}

We link intra–urban climate variability to urban characteristics via a simple logarithmic model as:
\begin{equation}
\Delta y_i \;=\; \alpha \;+\; \beta\,\ln x_i,
\label{eq:base_model}
\end{equation}
where \(\Delta y_i\) is the urban-rural climate difference in grid cell \(i\), $x$ the urban feature (e.g., street intersections), and \((\alpha,\beta)\) are city-level parameters. Taking the expected value and variance in equation \eqref{eq:base_model}, and replacing the city-specific intercepts $\alpha$ and slopes $\beta$ with cluster-specific averages $\overline{\alpha}_k$ and $\overline{\beta}_k$, where $k \in \{1,\ldots,K\}$ indexes clusters and $K$ is the total number of clusters, yields to:
\begin{equation}
    \langle \Delta y (x) \rangle = \overline{\alpha}_k + \overline{\beta}_k  \langle \ln x \rangle \quad \text{,} \quad \sigma_{\Delta y} (x) = \overline{\beta}_k \sigma_{\ln x}.
    \label{eq:si_log_mean_std_kmeans}
\end{equation}
The data collapse using the street network ($x{=}n$) as our predictor in equation \eqref{eq:si_log_mean_std_kmeans}, led to the optimal partition $K{=}3$ (quantified using the k-means clustering method; see SI Fig.~\ref{fig:SI_clustering_kmeans_popu_vs_node} and Methods for details). Examining the resulting parameter space, we find that even though the $(\alpha,\beta)$ plane does not perfectly separate the clusters, some clear patterns still emerge: the cluster-specific averages $(\overline{\alpha}_k,\overline{\beta}_k)$ differ significantly across clusters. Notably, Cluster $C_T$ and $C_{PM}$ (Purple) tend to have higher slope values, indicating that temperature and pollution levels in these cities are more sensitive to urban features (SI Table~\ref{table:SI_cluster_alpha_beta_absbeta}). We also observe that, for temperature (Fig.~\ref{fig:2}c), Cluster $A_T$ (blue) and Cluster $B_T$ (green) dominate much of Europe, the eastern United States, and Australia - regions characterized by milder temperate or humid-subtropical climates. In contrast, Cluster $C_T$ (purple) is prevalent across Africa and the western Americas (including Mexico, Chile, and Peru), consistent with hotter or more arid conditions. These temperature clusters therefore mirror differences in background climate, which is known to modulate the urban heat island effect \cite{Manoli2019}. For $PM$ (Fig.~\ref{fig:3}c), the Cluster $A_{PM}$ (Blue) and Cluster $B_{PM}$ (Green) prominently cover Europe, Japan, parts of the United States, and Australia, where there are relatively stricter pollution controls. Meanwhile, the Cluster $C_{PM}$ (Purple) includes high-emission areas in Asia, the Middle East, Africa, and South America, aligning with known pollution hot-spots and major industrial regions.

\subsection*{Clustering effect on the data collapse}
\begin{figure}[H]   
    \centering
    \includegraphics[width=\linewidth]{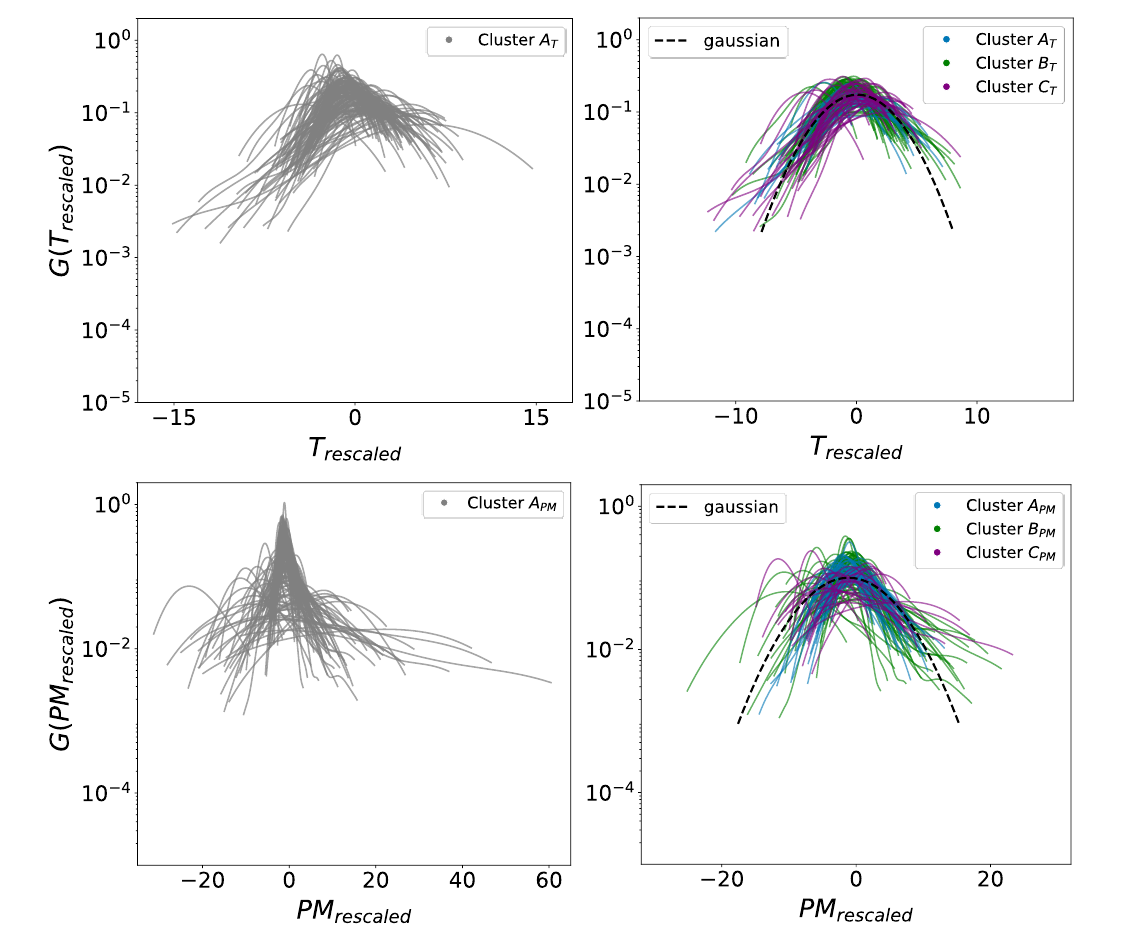} 
    \caption{\footnotesize Effect of $k$-means clustering on the data collapse when using the street network as the urban feature $x$ to predict urban climate PDFs, for $K=1$ (left) and $K=3$ (right). The left panel shows the data collapse using a global intercept $\overline{\alpha}$ and slope $\overline{\beta}$ values averaged across all 142 cities ($K=1$), while the right panel shows the collapse using group-specific average intercepts $\overline{\alpha}_k$ and slopes $\overline{\beta}_k$ for the optimal partition ($K=3$).}
\label{fig:SI_collapse_clustering}
\end{figure}

Here, we further explore how clustering improves the data collapse.
Assuming \(\ln x \sim \mathcal N\!\bigl(\langle \ln x \rangle,\sigma_{\ln x}^{2}\bigr)\) is approximately Gaussian shaped for urban features \cite{Hendrick2024}, the connection between the moments in the original space, $\langle x \rangle$ and $\sigma_x$, and those in log-space, $\langle \ln x \rangle$ and $\sigma_{\ln x}$ is
\begin{equation}
\langle x \rangle=e^{{\langle \ln x \rangle}+\frac12\sigma_{\ln x}^2},\qquad
\frac{\sigma_x^2}{\langle x \rangle^2}=e^{\sigma_{\ln x}^2}-1 
\end{equation}
and the inverses are
\begin{equation}
\sigma_{\ln x}^2 \;=\; \ln\!\big(1+\frac{\sigma_x^2}{\langle x \rangle^2}\big),\qquad
\langle ln x \rangle \;=\; \ln \langle x \rangle \;-\; \tfrac12 \ln\!\big(1+\frac{\sigma_x^2}{\langle x \rangle^2}\big).
\label{eq:mu_sigma_lny_from_y}
\end{equation}
Substituting equation \eqref{eq:mu_sigma_lny_from_y} into equation \eqref{eq:si_log_mean_std_kmeans} allows us to rewrite the mean and variance of the urban–rural climate difference as functions of the mean and variance of the urban feature in the original space (i.e. $\langle x \rangle$ and $\sigma_x$):
\begin{align}
\langle \Delta y \rangle
&= \overline{\alpha}_k \;+\; \overline{\beta}_k\left[\ln \langle x \rangle \;-\; \tfrac12 \ln\!\big(1+\frac{\sigma_x^2}{\langle x \rangle^2}\big)\right],
\label{eq:mean_in_y}
\\
\sigma_{\Delta y}
&= \overline{\beta}_k\,\sqrt{\,\ln\!\big(1+\frac{\sigma_x^2}{\langle x \rangle^2}\big)\,}.
\label{eq:sd_in_y}
\end{align}
Since $\langle x\rangle$ vs $\sigma_x$ are strongly correlated (linear fit, $R^2 \approx 0.8)$, the ratio $\sigma_x^2 / \langle x \rangle^2$ is nearly constant across cities. Consistently, we find that $\sigma_{\ln x}$ varies little across cities (in our data the coefficient of variation is $CV_{\sigma_{\ln x}} \simeq 0.18$). Therefore, because $\sigma_{\Delta y}=\overline{\beta}_k \sigma_{\ln x}$, cross-city variability in the rescaling factor is dominated by $\overline{\beta}_k$. Replacing a global $(\overline{\alpha},\overline{\beta})$ with cluster-specific $(\overline{\alpha}_k,\overline{\beta}_k)$ reduces cross-city variation in $\sigma_{\Delta y}$ and directly improves the data collapse under the location–scale ansatz (equation \ref{eq:ansatz}).
Intuitively, this means that cities differ mainly in how strongly their local climate responds to changes in urban structure, captured by the slope $\beta$. When we use a single global intercept $\overline{\alpha}$ and slope $\overline{\beta}$ (i.e., $K=1$), distinct families of curves emerge from the data collapse (see Fig.~\ref{fig:SI_collapse_clustering}). In other words, for $K=1$ we are under- or overestimating the characteristic scale of fluctuations for some cities. Clustering does not change the underlying scaling function $G$ — it simply aligns the location and scale across groups of cities that behave similarly, allowing all cities to collapse onto the same rescaled curve.

\section{Radial decay model of climate variables}
\label{sec:stochastic_decay_model}

The model developed by \cite{Zhou2015,Yu2020}, assuming a circular city of radius R and a radially decaying urban-rural climate difference variable, is given by:

\begin{equation}
\Delta y(r) \;=\; y_{A}\,\exp\!\Bigl(-\tfrac{r^{2}}{2\,\lambda_y^{2}}\Bigr),
\label{eq:decayModel_x}
\end{equation}
where $r$ denotes the radial distance from the city centre ($r=0$; see Methods for definitions), \(y_{A}\) is the climate central value at \(r=0\), \(\lambda_y\) controls the decay rate.

To derive the distribution of values \(\Delta y\), we note that a ring at radius \(r\) has area \(\mathrm{d}A = 2\pi r\, \mathrm{d}r\). Since \(\Delta y\) decreases with \(r\), the probability density \(\rho(\Delta y)\,\mathrm{d}\Delta y\) is proportional to the fraction of area between \(r\) and \(r + \mathrm{d}r\), yielding:

\begin{equation}
\rho(\Delta y)\,\mathrm{d}\Delta y
\;=\;
\frac{dA}{A}
\;=\;
\frac{2\,r}{R^{2}}
\,\bigl|\tfrac{\mathrm{d}r}{\mathrm{d}\Delta y}\bigr|
\,\mathrm{d}\Delta y
\quad\Longrightarrow\quad
\rho(\Delta y)
=
\frac{2\,r}{R^{2}}
\,\bigl|\tfrac{\mathrm{d}r}{\mathrm{d}\Delta y}\bigr|.
\end{equation}
From equation \eqref{eq:decayModel_x}, we can write \(\ln(\Delta y/y_{A})=-\tfrac{r^{2}}{2\lambda_y^{2}}\), that gives
\begin{equation}
r
=
\lambda_y\,\sqrt{-\,2\,\ln\!\Bigl(\tfrac{\Delta y}{y_{A}}\Bigr)}
\quad
\text{and}
\quad
\frac{\mathrm{d}r}{\mathrm{d}\Delta y}
=
-\frac{\lambda_y^{2}}{r\,\Delta y}.
\end{equation}
All together, we obtain:
\begin{equation}
\rho(\Delta y)
=
\frac{2\,\lambda_y^{2}}{R^{2}}
\,\frac{1}{\Delta y} \propto 1/\Delta y\,
\end{equation}
that is valid for \(\Delta y \in \big[\,y_{A}\,e^{-\,R^{2}/(2\,\lambda_y^{2})},\,y_{A}\big]\). One can verify that \(\int \rho(\Delta y)\,\mathrm{d}\Delta y = 1\), so we obtain \(\rho(\Delta y) \propto 1/\Delta y\).

\vspace{1em}
\noindent
\subsection*{Adding additive white noise} 
Empirical data exhibit fluctuations around the deterministic decay in equation \eqref{eq:decayModel_x}. To capture these fluctuations, we add an additive noise term:

\begin{equation}
\Delta y(r)
\;=\;
y_{A} \exp \Bigl(-\tfrac{r^{2}}{2\,\lambda_y^{2}}\Bigr)
\;+\;
\mathcal{N}\bigl(0,\,\sigma_{r,\text{city}}^{2}\bigr).
\end{equation}
Here, the noise strength \(\sigma_{r,\text{city}}\) has the same units as \(\Delta y\). This white noise convolves the original deterministic distribution \(\rho(\Delta y) \propto 1/\Delta y\) with a Gaussian kernel. When \(\sigma_{r,\text{city}}\) is small, the \(1/\Delta y\) shape is only slightly blurred; for large noise, the distribution may significantly deviate from \(1/\Delta y\), approaching a Gaussian — matching what is observed in empirical climate data. 
It is interesting to note that, when normalized by their central amplitudes $y_A$, the noise strengths of temperature and pollution exhibit a moderate positive Pearson correlation ($C=0.34$).

\vspace{1em}

\noindent
\subsection*{Linking climate radial decay model to urban features}
For an urban feature $x$ (e.g., street intersections) we use the same deterministic form:

\begin{equation}
x(r) \;=\; x_{A}\,\exp\!\Bigl(-\tfrac{r^{2}}{2\,\lambda_x^{2}}\Bigr),
\label{eq:decayModel_y}
\end{equation}
where $r$ denotes the radial distance from the city centre ($r=0$; see Methods for definitions), \(x_{A}\) is the central value at \(r=0\), and \(\lambda_x\) controls the decay rate of the urban feature. Repeating the above derivations gives $\rho(x) \propto 1/x$.

To better reproduce the empirical distribution of urban features, we follow the same approach used for climate variables, but now fluctuations are multiplicative rather than additive. Thus, we add a multiplicative white noise to equation \ref{eq:decayModel_y} as:

\begin{equation}
    x(r)= x_{A}\,\exp\!\Bigl(-\tfrac{r^{2}}{2\,\lambda_x^{2}}\Bigr) e^{\epsilon} \quad \text{with} \quad \epsilon\sim \mathcal N\bigl(0,\sigma_{r,\text{city}}^{x}\,^{2}\bigr)
\end{equation}
or, equivalently:

\begin{equation}
\ln x(r)\;=\;
  \Bigl[\ln x_{A}-\frac{r^{2}}{2\lambda_x^{2}}\Bigr]
  \;+\;
    N\bigl(0,\sigma_{r,\text{city}}^{x}\,^{2}\bigr)
\label{eq:log_noise}
\end{equation}
where the noise strength $\sigma^{x}_{r,\text{city}}$ is dimensionless. This white noise convolves the deterministic distribution $\rho(\ln x)$ with a Gaussian kernel. As the noise strength increases, the resulting distribution becomes Gaussian: 

\begin{equation}
    \rho(\ln x) \sim \text{gaussian} \quad\Longrightarrow\quad \rho(x) \sim \text{log-normal}
\end{equation}
reproducing the empirical findings of log-normal PDFs for urban features \cite{Hendrick2024}.

\section{Validation with alternative datasets}
\label{sec:alternative_datasets}

To further assess the robustness of our results, we validated the scaling functions using independent methodological datasets for both climate variables (see Methods in the main text for details).

\begin{figure}[H]   
    \centering
    \includegraphics[width=\linewidth]{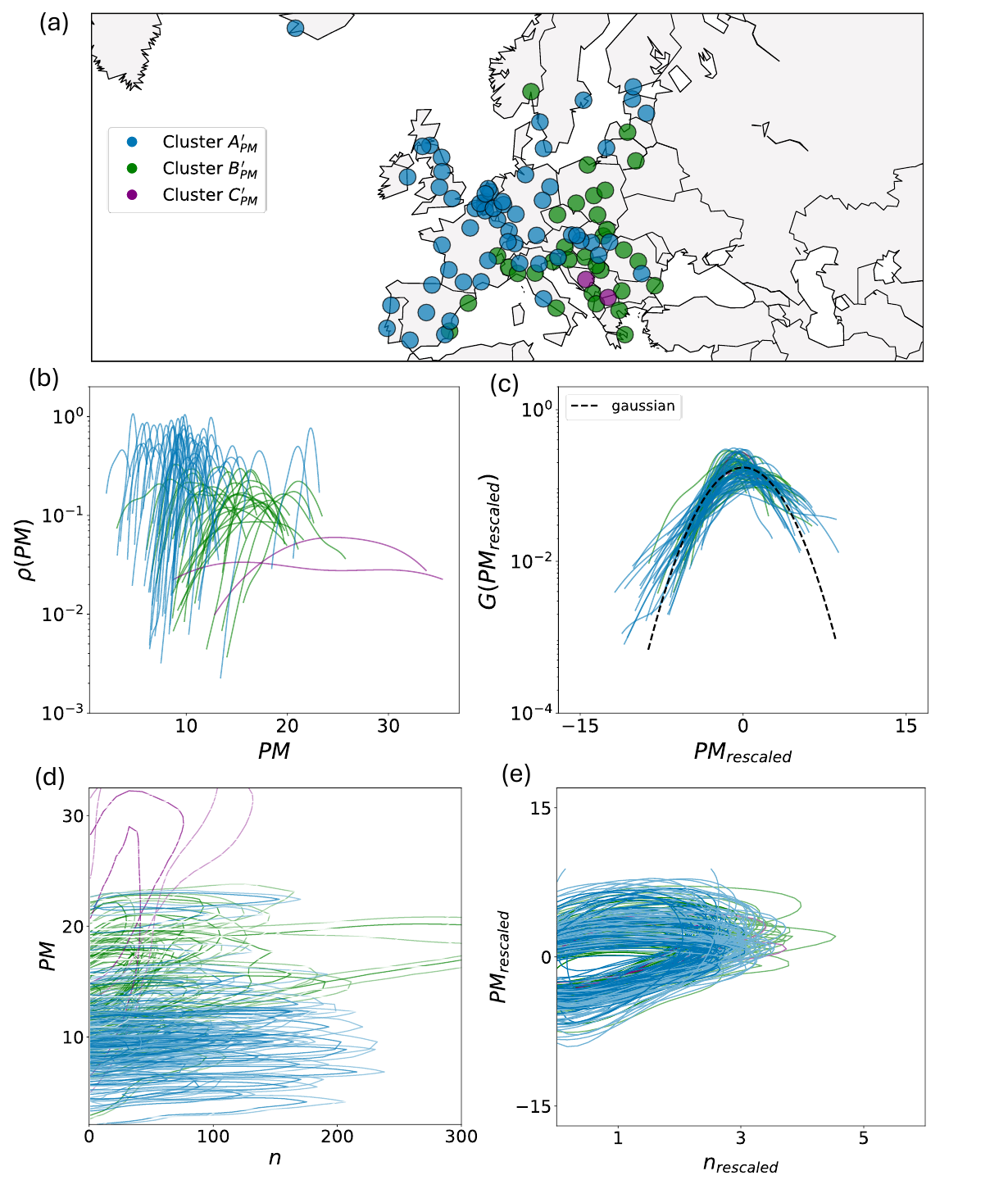} 
    \caption{\footnotesize Confirming the robustness of the underlying scaling framework using a regression-interpolation framework from pollution ground-based monitoring stations in Europe during 2022.}
\label{fig:SI_double_check_PM}
\end{figure}

First, we analyzed the PM results using an independent European dataset at a 1 km resolution for the same year \cite{pm25data_europe}. While our primary global dataset relies on a machine-learning approach integrating satellite aerosol optical depth retrievals, ground-based monitoring stations, and meteorological data, the European dataset employs a regression-interpolation framework that spatially extrapolates PM values from monitoring networks. Despite methodological differences, the scaling functions remain valid, confirming the robustness of our scaling framework (Fig. \ref{fig:SI_double_check_PM}). Furthermore, the clustering structure remains largely consistent across datasets, reinforcing the stability of the observed patterns. Both datasets reveal two dominant PM clusters in Europe, with the higher-resolution European dataset enhancing the separation between eastern and western European cities. This indicates that the broad-scale PM patterns observed globally persist across resolutions and methodologies, highlighting the resilience of the scaling framework.

\begin{figure}[H]   
    \centering
    \includegraphics[width=\linewidth]{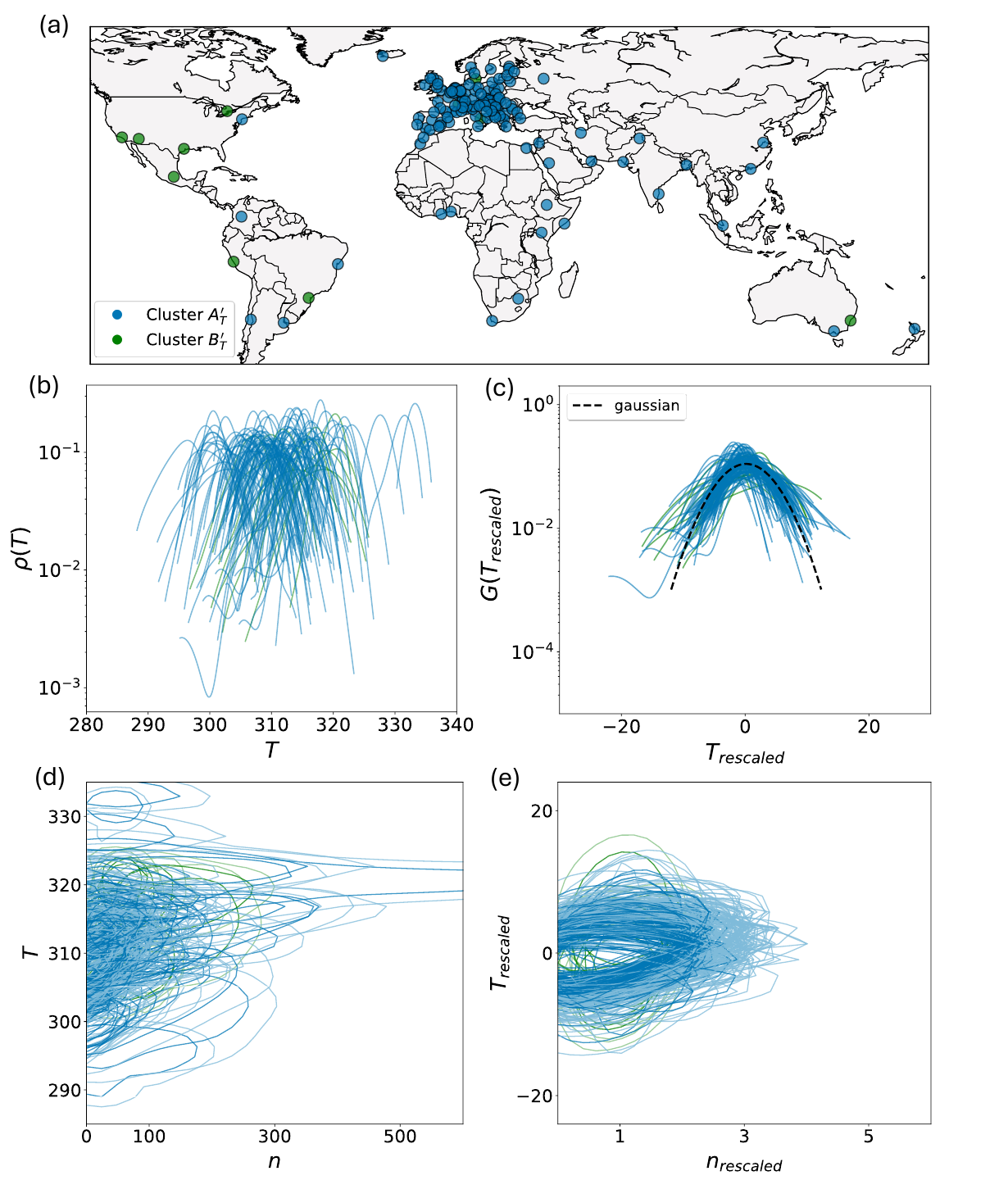} 
    \caption{\footnotesize Confirming the robustness of the underlying scaling framework using observational maximum land surface temperature (LST) data during Summer months 2013.}
\label{fig:SI_double_check_T}
\end{figure}

Similarly, we validated the consistency of our temperature results using an independent observational dataset of land surface temperature (LST) at 1 km resolution \cite{temperature_observational}. The datasets correspond to different years and differ methodologically: our primary dataset is based on high-resolution urban climate modeling estimating near-surface air temperature, while the comparison dataset provides satellite-derived land surface temperature (Tmax), representing the highest recorded daytime temperatures during summer months. Despite these methodological differences, the scaling functions remain valid in both cases (Fig.~\ref{fig:SI_double_check_T}). However, clustering analyses reveal differing behaviors. Specifically, summer land surface temperature (Tmax) exhibits a more compressed Gaussian distribution, resulting in limited clustering differentiation compared to the wider distribution of annual air mean temperatures, which better discriminates distinct climatic regimes. This suggests that the yearly air mean temperature captures a broader spectrum of intra-urban climate variability, enabling clearer climatic differentiation across cities.

Thus, despite methodological variations, our scaling framework remains robust for both climate variables, further demonstrating its general applicability.

\section{Multivariate extension of the logarithmic model}
\label{sec:multiloglinear}
The logarithmic model in equation \eqref{eq:log} can be extended to a multivariate model in which each climate variable $\Delta y_j$ (e.g., temperature or particulate matter) depends on several urban features $\{x_i\}$ (e.g., street intersections, population, vegetation), according to:
\begin{equation}
  \Delta y_j = \alpha_j + \sum_i \beta_{ji} \ln x_i.
  \label{eq:multiloglinear}
\end{equation}
This formulation allows for a multidimensional relationship between climate and urban structure, where each coefficient $\beta_{ji}$ quantifies the sensitivity of the climate variable $\Delta y_j$ to changes in a given urban feature $x_i$. While such a multivariate approach would likely increase explanatory power, it would also introduce additional free parameters and potential correlations between covariates, thereby reducing interpretability and generality across cities. In our study, we adapt a simple single model since our aim here is simplicity and generality to test whether a general scaling function exists for the intra-urban distributions of temperature and pollution. Future studies could leverage it to examine how multiple urban features jointly determine local climate variability at high spatial resolution, particularly in contexts where fine-scale predictability is the primary objective.

\vspace{1em}

\newpage
\section*{Supplementary Figures and Tables}

\begin{table}[H]
\centering
\renewcommand{\arraystretch}{1.3}
\footnotesize
\caption{Cities included in the study.}
\vspace{1em}
\label{table:SI_cities}

\begin{tabular}{llll}
\multicolumn{4}{c}{\textbf{Cities}} \\
\midrule
Tirana, Albania & Buenos Aires, Argentina & Melbourne, Australia & Sydney, Australia \\
Graz, Austria & Vienna, Austria & Dhaka, Bangladesh & Antwerp, Belgium \\
Brussels, Belgium & Charleroi, Belgium & Ghent, Belgium & Liège, Belgium \\
Sarajevo, Bosnia and Herzegovina & Curitiba, Brazil & Salvador, Brazil & Sofia, Bulgaria \\
Varna, Bulgaria & Toronto, Canada & Santiago, Chile & Hong Kong, China \\
Nanjing, China & Bogotá, Colombia & Split, Croatia & Zagreb, Croatia \\
Prague, Czechia & Copenhagen, Denmark & Cairo, Egypt & Tallinn, Estonia \\
Tartu, Estonia & Addis Ababa, Ethiopia & Helsinki, Finland & Bordeaux, France \\
Lille, France & Lyon, France & Marseille, France & Montpellier, France \\
Nantes, France & Nice, France & Paris, France & Strasbourg, France \\
Toulouse, France & Berlin, Germany & Cologne, Germany & Düsseldorf, Germany \\
Frankfurt, Germany & Hamburg, Germany & Leipzig, Germany & Munich, Germany \\
Accra, Ghana & Athens, Greece & Thessaloniki, Greece & Budapest, Hungary \\
Debrecen, Hungary & Győr, Hungary & Miskolc, Hungary & Pécs, Hungary \\
Szeged, Hungary & Reykjavík, Iceland & Chennai, India & Jakarta, Indonesia \\
Tehran, Iran & Dublin, Ireland & Bari, Italy & Bologna, Italy \\
Genoa, Italy & Milan, Italy & Naples, Italy & Padua, Italy \\
Palermo, Italy & Rome, Italy & Trieste, Italy & Turin, Italy \\
Tokyo, Japan & Amman, Jordan & Nairobi, Kenya & Riga, Latvia \\
Klaipėda, Lithuania & Vilnius, Lithuania & Luxembourg, Luxembourg & Mexico City, Mexico \\
Podgorica, Montenegro & Marrakesh, Morocco & Rabat, Morocco & Amsterdam, Netherlands \\
Rotterdam, Netherlands & Utrecht, Netherlands & Auckland, New Zealand & Lagos, Nigeria \\
Skopje, North Macedonia & Oslo, Norway & Islamabad, Pakistan & Karachi, Pakistan \\
Lima, Peru & Gdańsk, Poland & Kraków, Poland & Warsaw, Poland \\
Wrocław, Poland & Łódź, Poland & Lisbon, Portugal & Porto, Portugal \\
Brașov, Romania & Bucharest, Romania & Cluj-Napoca, Romania & Moscow, Russia \\
Medina, Saudi Arabia & Dakar, Senegal & Belgrade, Serbia & Novi Sad, Serbia \\
Singapore, Singapore & Bratislava, Slovakia & Košice, Slovakia & Ljubljana, Slovenia \\
Mogadishu, Somalia & Cape Town, South Africa & Tshwane, South Africa & Alicante, Spain \\
Barcelona, Spain & Bilbao, Spain & Madrid, Spain & Murcia, Spain \\
Málaga, Spain & Palma de Mallorca, Spain & Seville, Spain & Valencia, Spain \\
Gothenburg, Sweden & Stockholm, Sweden & Basel, Switzerland & Geneva, Switzerland \\
Zurich, Switzerland & Istanbul, Türkiye & Dubai, United Arab Emirates & Birmingham, United Kingdom \\
Edinburgh, United Kingdom & Glasgow, United Kingdom & Leeds, United Kingdom & London, United Kingdom \\
Newcastle, United Kingdom & Houston, United States & Los Angeles, United States & New York, United States \\
Phoenix, United States & Ho Chi Minh City, Vietnam &  &  \\
\end{tabular}
\end{table}

\vspace{1em}

\begin{table}[H]
\centering
\caption{Numerical comparison of logarithmic, linear, and power-law models between intra-urban climate variables and urban features. The $R^2$ is the coefficient of determination; RMSE is the root mean squared error; and $p_{\mathrm{BP}}$ is the two-sided $p$-value of the Breusch–Pagan test for heteroscedasticity of residuals. Values represent the mean~$\pm$~standard deviation across all 142 cities.}
\vspace{1em}

\textbf{(A) Street intersections ($n$)}\\[0.3em]
\begin{tabular}{lccc ccc}
 & \multicolumn{3}{c}{Temperature ($\Delta T$)} & \multicolumn{3}{c}{Pollution ($\Delta PM$)} \\
 & R$^2$ & RMSE & $p_{\mathrm{BP}}$ & R$^2$ & RMSE & $p_{\mathrm{BP}}$ \\
\cmidrule(lr){2-4} \cmidrule(lr){5-7}
Linear & 0.33 ± 0.16 & 0.54 ± 0.29 & 0.19 ± 0.27 & 0.17 ± 0.14 & 1.40 ± 1.35 & 0.16 ± 0.27\\
Log--linear & 0.33 ± 0.14 & 0.53 ± 0.27 & 0.14 ± 0.24 & 0.13 ± 0.11 & 1.43 ± 1.37 & 0.14 ± 0.24\\
Power--law & 0.36 ± 0.16 & 0.52 ± 0.28 & 0.17 ± 0.27 & 0.17 ± 0.14 & 1.40 ± 1.34 & 0.15 ± 0.26\\
\end{tabular}

\vspace{1em}

\textbf{(B) Population counts ($p$)}\\[0.3em]
\begin{tabular}{lccc ccc}
 & \multicolumn{3}{c}{Temperature ($\Delta T$)} & \multicolumn{3}{c}{Pollution ($\Delta PM$)}\\
 & R$^2$ & RMSE & $p_{\mathrm{BP}}$ & R$^2$ & RMSE & $p_{\mathrm{BP}}$\\
\cmidrule(lr){2-4} \cmidrule(lr){5-7}
Linear & 0.42 ± 0.20 & 0.49 ± 0.29 & 0.11 ± 0.23 & 0.24 ± 0.19 & 1.31 ± 1.21 & 0.13 ± 0.27\\
Log--linear & 0.48 ± 0.22 & 0.47 ± 0.27 & 0.10 ± 0.22 & 0.23 ± 0.16 & 1.33 ± 1.21 & 0.13 ± 0.27\\
Power--law & 0.47 ± 0.21 & 0.47 ± 0.28 & 0.09 ± 0.21 & 0.25 ± 0.20 & 1.30 ± 1.20 & 0.15 ± 0.27\\
\end{tabular}

\label{table:SI_fit_models}
\end{table}

\vspace{1em}

\begin{table}[H]
\centering
\caption{Cluster summaries ($K=3$) of the logarithmic model parameters obtained from $k$-means clustering using street intersections ($n$). For each cluster we report sample sizes ($n_T$, $n_{PM}$) and the cluster-specific averages: intercept $\overline{\alpha}_k$ , slope $\overline{\beta}_k$, and absolute slope $|\overline{\beta}_k|$ across cities. Values are reported as mean~$\pm$~standard deviation [95\% confidence interval] across all 142 cities.}

\vspace{1em}

\textbf{(A) Temperature}\\[0.3em]
\begin{tabular}{l c c c c}
Cluster & $n_T$ & $\overline{\alpha}_k$ & $\overline{\beta}_k$ & $|\overline{\beta}_k|$\\
\cmidrule(lr){2-5}
$A_T$ & 18 & $-0.14 \pm 0.15$ [$-0.21$, $-0.06$] & $0.11 \pm 0.06$ [0.09, 0.14] & $0.11 \pm 0.05$ [0.09, 0.14] \\
$B_T$ & 86 & $-0.20 \pm 0.31$ [$-0.27$, $-0.14$] & $0.26 \pm 0.10$ [0.24, 0.28] & $0.26 \pm 0.10$ [0.24, 0.28] \\
$C_T$ & 38 & $-0.23 \pm 0.52$ [$-0.40$, $-0.06$] & $0.39 \pm 0.12$ [0.35, 0.43] & $0.39 \pm 0.12$ [0.35, 0.43] \\
\end{tabular}

\vspace{1em}

\textbf{(B) Pollution}\\[0.3em]
\begin{tabular}{l c c c c}
Cluster & $n_{PM}$ & $\overline{\alpha}_k$ & $\overline{\beta}_k$ & $|\overline{\beta}_k|$\\
\cmidrule(lr){2-5}
$A_{PM}$ & 42 & $-0.08 \pm 0.11$ [$-0.11$, $-0.05$] & $0.10 \pm 0.08$ [0.07, 0.12] & $0.11 \pm 0.07$ [0.08, 0.13] \\
$B_{PM}$ & 76 & $-0.32 \pm 0.55$ [$-0.45$, $-0.20$] & $0.30 \pm 0.33$ [0.22, 0.38] & $0.37 \pm 0.26$ [0.31, 0.43] \\
$C_{PM}$ & 24 & $-1.05 \pm 2.90$ [$-2.27$, 0.18] & $0.82 \pm 1.05$ [0.37, 1.26] & $1.00 \pm 0.88$ [0.63, 1.37] \\
\end{tabular}

\label{table:SI_cluster_alpha_beta_absbeta}
\end{table}

\vspace{1em}

\begin{table}[H]
\centering
\renewcommand{\arraystretch}{1.8}
\caption{Pearson correlation coefficients ($C$) between decay rates ($\lambda$) and stochastic noise strengths ($\sigma_{r,\mathrm{city}}$) of urban structure (street network $n$, population $p$) and climate variables (temperature $T$, pollution $PM$) in the stochastic decay model across all cities included for the spatial decay analysis ($n = 107$ for temperature and $n = 95$ for $PM$; see Methods). Statistical significance was assessed using two-sided tests; exact $p$-values are reported.}
\vspace{1em}

\label{table:SI_correlations}
\setlength{\tabcolsep}{12pt}
\begin{tabular}{lcc}
 & C (street network, $n$) & C (population, $p$) \\
\cmidrule(lr){2-3}
$\lambda_{n,p}$ vs.\ $\lambda_T$ 
& 0.80 ($p = 4.3 \times 10^{-25}$) 
& 0.57 ($p = 2.6 \times 10^{-10}$) \\

$\lambda_{n,p}$ vs.\ $\lambda_{PM}$ 
& 0.55 ($p = 4.9 \times 10^{-9}$) 
& 0.19 ($p = 6.6 \times 10^{-2}$) \\

$\sigma^{n,p}_{r,\mathrm{city}}$ vs.\ $\sigma^{T}_{r,\mathrm{city}}$ 
& 0.39 ($p = 4.1 \times 10^{-5}$) 
& 0.14 ($p = 0.16$) \\

$\sigma^{n,p}_{r,\mathrm{city}}$ vs.\ $\sigma^{PM}_{r,\mathrm{city}}$ 
& 0.05 ($p = 0.62$) 
& 0.08 ($p = 0.45$) \\
\end{tabular}
\end{table}

\newpage



\begin{figure}[H]   
    \centering
    \includegraphics[width=\linewidth]{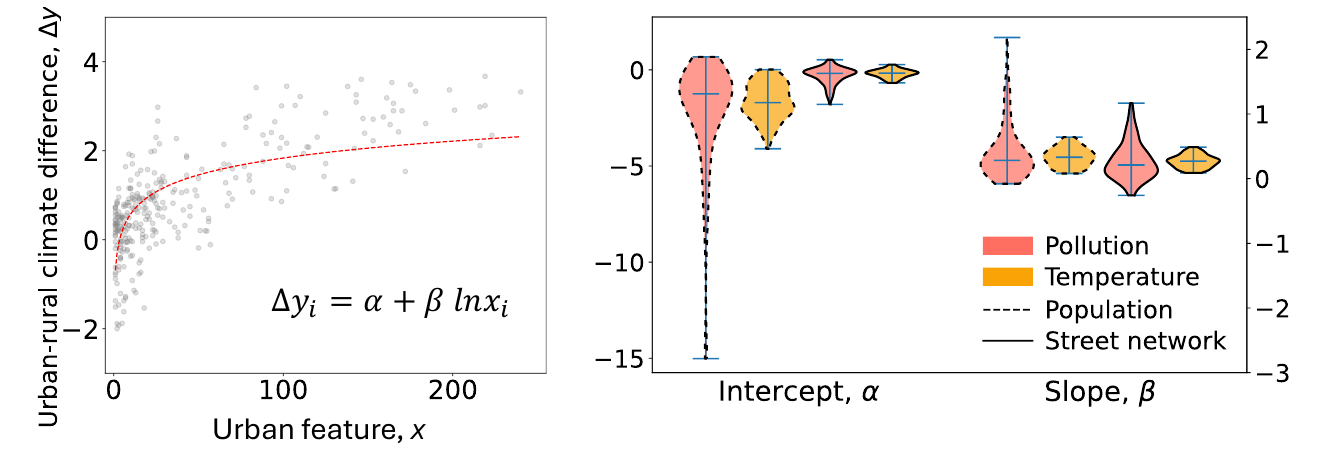} 
    \caption{\footnotesize \textbf{Logarithmic scaling relationship between intra-urban climate variability and urban features.} Comparative analysis of the intercept ($\alpha$) and the slope ($\beta$) of the logarithmic relationship between intra‐urban climate variability and urban features across all 142 cities. Violin plots display the central 95\% of the parameters estimates, highlighting that street network is more consistent and robust than population counts.}
\label{fig:SI_log_results}
\end{figure}


\begin{figure}[H]   
    \centering
    \includegraphics[width=\linewidth]{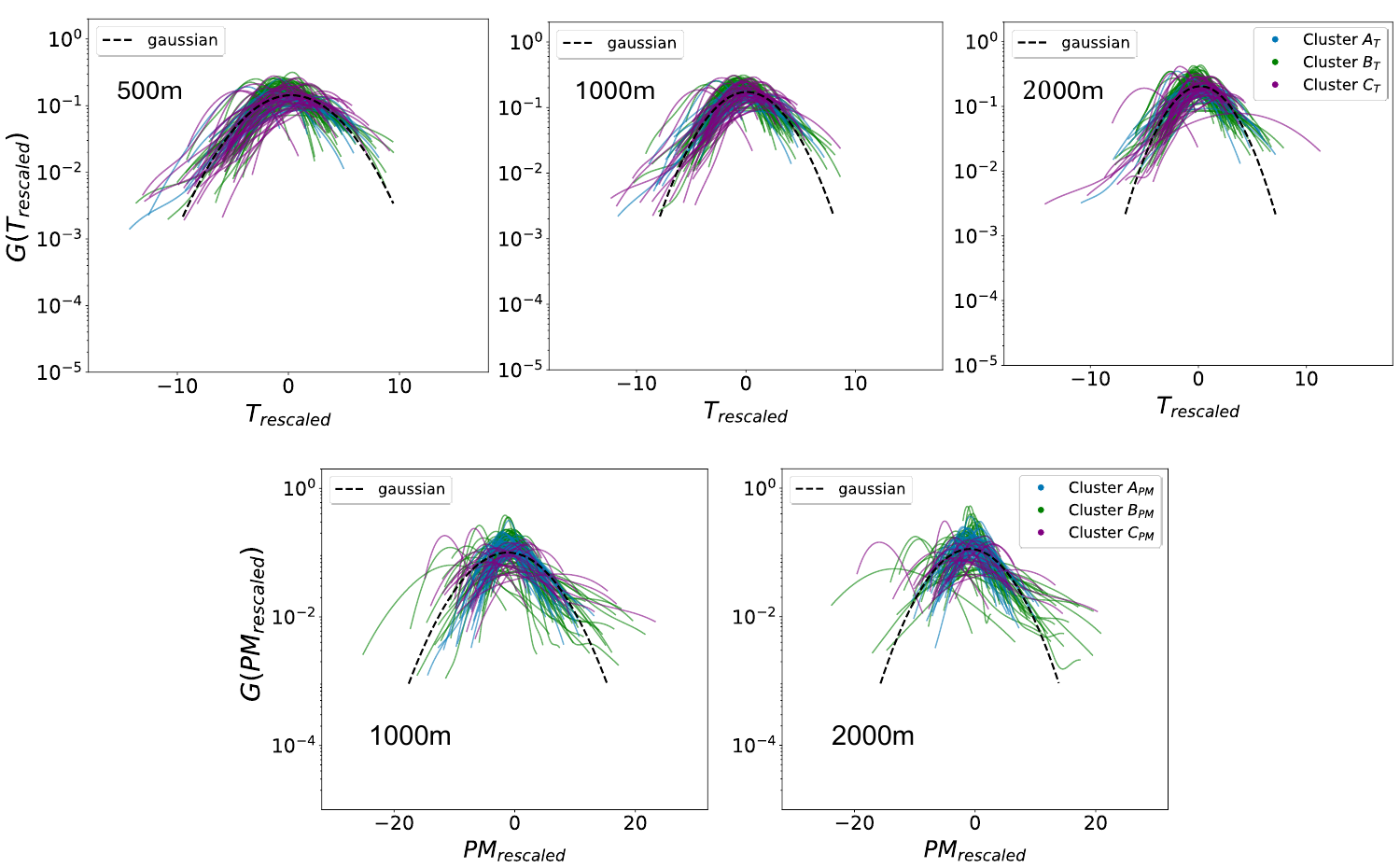} 
    \caption{\footnotesize The results remain robust across different spatial scales of the regular square grid, where each square cell has a side length size $l=500m$,$1000m$,$2000m$.}
\label{fig:SI_scales}
\end{figure}

\begin{figure}[H]   
    \centering
    \includegraphics[width=0.85\textwidth]{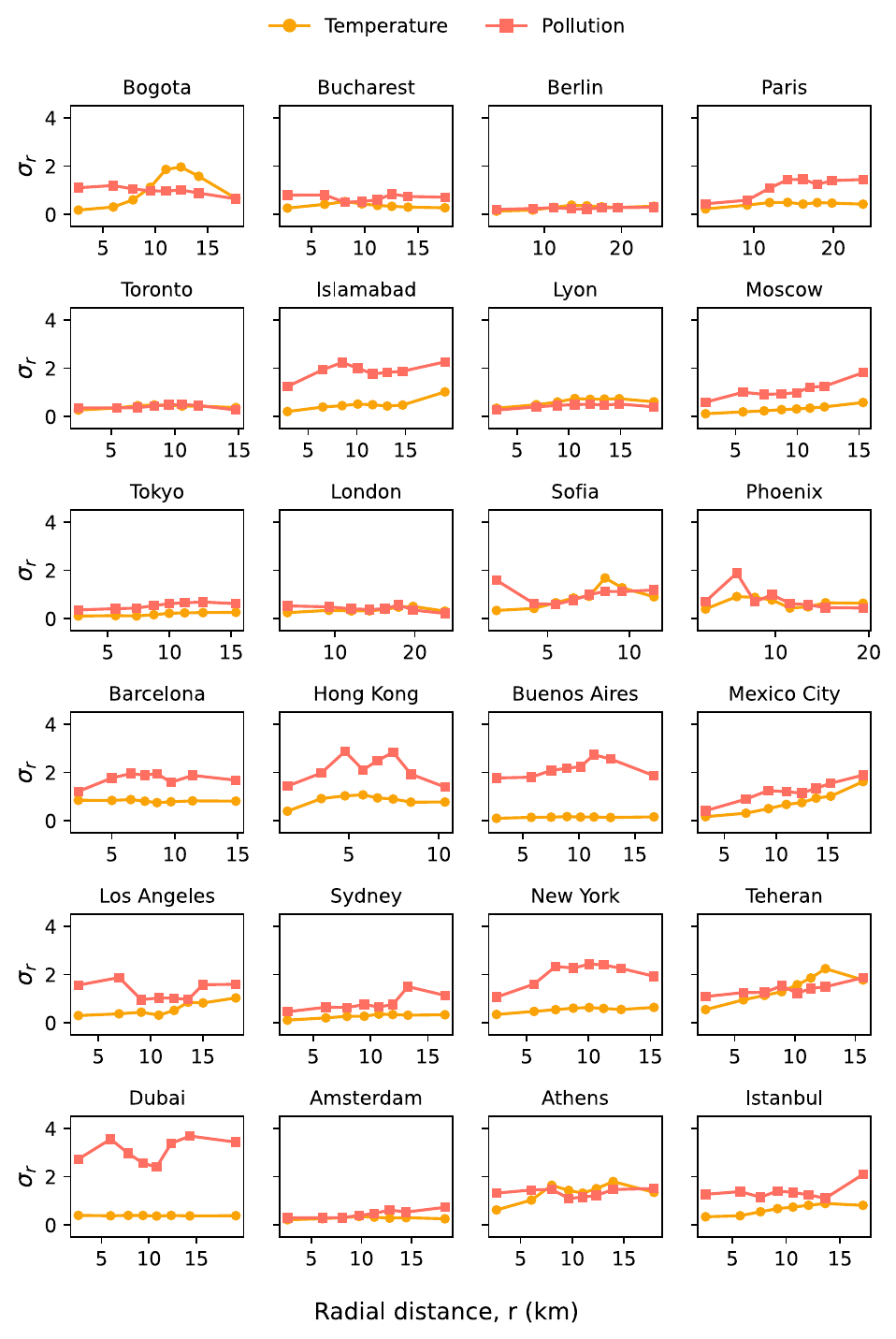}
    \caption{\footnotesize \textbf{Radial distance dependence of residual from the mean decay profile.} We compute the empirical noise strength ($\sigma_r$) of temperature and particulate matter within concentric belts as the standard deviation of residuals relative to the fitted deterministic decay. Across cities of different climate and scale, $\sigma_r$ remains approximately constant with radial distance. This supports modeling the stochastic component as additive white noise with a single city-level variance parameter ${\sigma_{r,\text{city}}}^{2}$.}
\label{fig:SI_illustration_decay_cities_noise}
\end{figure}


\begin{figure}[H]   
    \centering
    \includegraphics[width=\linewidth]{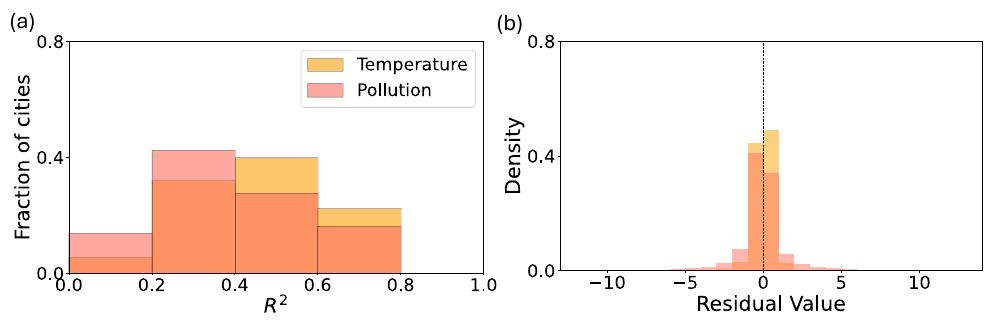} 
        \caption{\footnotesize \textbf{Radial decay model performance for climate variables.} (a) Shows the $R^2$ values across all cities included for the spatial decay analysis for both temperature and particulate matter concentrations and, (b) the histogram of all aggregated city residuals.}
\label{fig:SI_decay_results}
\end{figure}

\begin{figure}[H]   
    \centering
    \includegraphics[width=\linewidth]{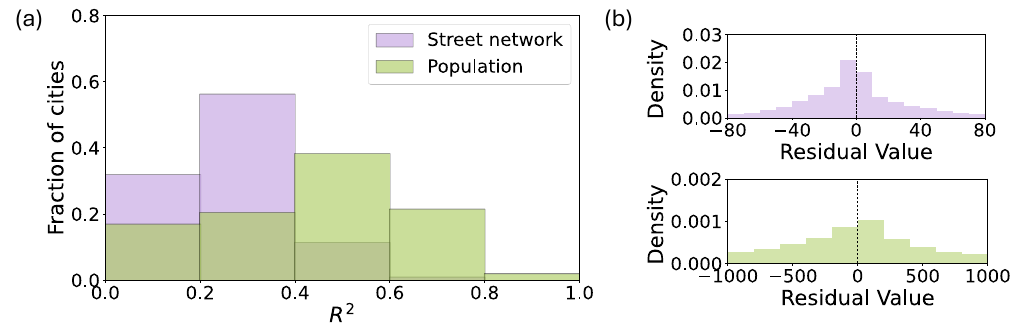} 
        \caption{\footnotesize \textbf{Radial decay model performance for urban features.} (a) Shows the $R^2$ values across all cities included for the spatial decay analysis. (b) the histogram of all aggregated city residuals. }
\label{fig:SI_decay_results_node}
\end{figure}

\begin{figure}[H]   
    \centering
    \includegraphics[width=\linewidth]{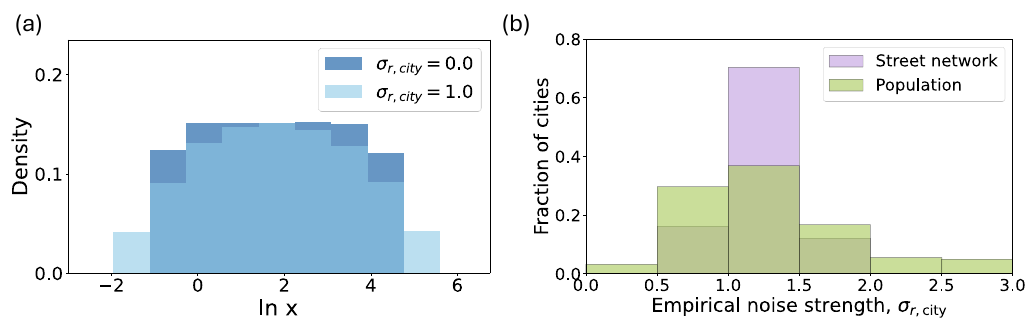} 
        \caption{\footnotesize \textbf{Stochastic decay model for urban features.} (a) Probability density functions from numerical simulations: when $\sigma_r^{city}=0.0$, the deterministic decay model is recovered, while for $\sigma_r^{city}=1.0$ the distribution of $\ln x$ approaches a Gaussian shape. The simulations use a peak value $x_A = 100$, decay rate $\lambda_x = 7.5\ \text{km}$, and city radius $R = 25\ \text{km}$. (b) Empirical distribution of noise strengths ($\sigma_r^{city}$) estimated from street network intersection and population counts data across all cities included for the spatial decay analysis.}
\label{fig:SI_stochastic_decay_results_node}
\end{figure}




\end{document}